\documentclass{aa}

\usepackage{graphics}

\def\ie{{\it i.e.}}
\def\ki2{$\chi^2$}
\def\Deltaki2{$\Delta\chi^2$}

\def\lya{Lyman~$\alpha$}
\def\cm2{cm$^{-2}$}
\def\cmm3{cm$^{-3}$}
\def\dsh{D/H}
\def\dshprim{(D/H)$_{prim}$}

\def\dshism{(D/H)$_{ISM}$}
\def\dshlic{(D/H)$_{\mathrm{LIC}}$}
\def\dshbc{(D/H)$_{\mathrm{BC}}$}
\def\kms{${\rm km\,s}^{-1}$}
\def\eg{{\it e.g.}}

\begin{document}

   \thesaurus{08     		 % A&A Section 8: Diffuse matter in space
              (08.09.2 Sirius A; % Stars: individual: Sirius A,
               08.09.2 Sirius B; % Stars: individual: Sirius B,
               08.23.1; 	 % Stars: white dwarfs,
               09.01.1; 	 % ISM: abundances,
               13.21.3; 	 % Ultraviolet: ISM,
               13.21.5)}	 % Ultraviolet: stars.

  \title{Ultraviolet observations of Sirius~A and Sirius~B with HST-GHRS
		\thanks{Based on observations with the NASA/ESA
               {\em Hubble Space Telescope}, obtained at the Space
               Telescope Science Institute, which is operated
               by the Association of Universities for Research in 
               Astronomy, Inc. under NASA contract No.~NAS5-26555.}}

  \subtitle{An interstellar cloud with a possible low deuterium abundance}

  \author{G.~H\'ebrard \inst{1, 2}
	  \and
		C.~Mallouris \inst{3}
	  \and
		R.~Ferlet \inst{1}
	  \and
		D.~Koester \inst{4}
	  \and
		M.~Lemoine \inst{5}
	  \and
		A.~Vidal-Madjar \inst{1}
	  \and
		D.~York \inst{3}
          }

   \offprints{Guillaume H\'ebrard}
   \mail{hebrard@iap.fr}

   \institute{Institut d'Astrophysique de Paris, CNRS,
              	98 bis Boulevard Arago, F-75014 Paris, France
         \and
                Department of Physics and Astronomy, 
                    Johns Hopkins University, 
                    3400 North Charles Street, Baltimore, MD~21218, USA
         \and
		University of Chicago, Department of Astronomy and 
		Astrophysics, 5640 South Ellis Avenue, Chicago, 
                IL~60637, USA
	 \and
		Institut f\"ur Theoretische Physik und Astrophysik der 
                     Christian-Albrechts-Universit\"at, D-24098 Kiel, 
                     Germany
	 \and
		DARC, UPR-176 CNRS, Observatoire de Paris-Meudon, 
			F-92195 Meudon C\'edex, France
             }

   \date{Received ? / Accepted ?}

   \maketitle

\begin{abstract}

We present new observations of the binary Sirius~A~/ Sirius~B performed 
with HST-GHRS. Two interstellar clouds are detected on this sightline, 
one of them being identified as the Local Interstellar Cloud (LIC), in 
agreement with previous HST-GHRS observations of Sirius~A 
(Lallement et al.~\cite{lalle}). The interstellar structure of this 
sightline, which we assume is the same toward both stars 
(separated by less than 4~arcsec at the time of observation), is 
constrained by high spectral resolution data of the species  
\ion{O}{i}, \ion{N}{i}, \ion{Si}{ii}, \ion{C}{ii}, \ion{Fe}{ii} and 
\ion{Mg}{ii}. 

\lya\ interstellar lines are also observed toward the two stars. But 
whereas the deuterium \lya\ line is well detected in the LIC with an 
abundance in agreement with that obtained by 
Linsky et al.~(\cite{linsky93} \&~\cite{linsky95}), no significant 
\ion{D}{i} line is detected in the other cloud. 

However, the \lya\ lines toward Sirius~A and Sirius~B are not trivial.
An excess of absorption is seen in the blue wing of the Sirius~A
\lya\ line and interpreted as the wind from Sirius~A. In its white
dwarf companion, an excess in absorption is seen in the red wing and
interpreted as the core of the Sirius~B photospheric \lya\ line. 
A composite \lya\ profile can nonetheless be constructed, and 
allows one to measure the deuterium abundance in the second cloud
$0<$\dshism$<1.6\times10^{-5}$, which is marginally in agreement with 
the Linsky et al.~(\cite{linsky93} \&~\cite{linsky95}) value. This 
sightline appears consequently as a good candidate for a low \dshism.

      \keywords{Stars: individual: Sirius A -- 
Stars: individual: Sirius B -- Stars: white dwarfs -- ISM: abundances -- 
Ultraviolet: ISM -- Ultraviolet: stars.}
   \end{abstract}

\section{Introduction}
\label{intro}

It is generally believed that deuterium is only produced in primordial
Big Bang nucleosynthesis (BBN), and destroyed in stellar interiors
(Epstein, Lattimer \& Schramm~\cite{epstein76}).  Hence, any
abundance of deuterium measured at any metallicity should provide a
lower limit to the primordial deuterium abundance 
(Reeves et al.~\cite{reeves73}). Deuterium is thus a key element in 
cosmology and in galactic chemical evolution 
(\eg,~Vangioni-Flam \& Cass\'e~\cite{flam95}; Prantzos~\cite{prantzos96}; 
Scully et al.~\cite{scully97}).  The primordial abundance of deuterium 
is indeed one of the best probes of the baryonic density parameter of 
the Universe $\Omega_B$. The decrease of its abundance all along 
galactic evolution, amongst other things, is a function of the star 
formation rate. Standard models predicting a decrease by a factor 2 to 3 
in 15~Gyrs (\eg,~Galli et al.~\cite{galli95}; Prantzos~\cite{prantzos95}; 
Tosi et al.~\cite{tosi98}). However there are some nonstandard models 
which propose nonprimordial deuterium production 
(see~\eg\ Lemoine et al.~(\cite{lemoine99}) for a~review). 
The most recent paper on nonprimordial deuterium production 
is by Mullan \& Linsky~(\cite{mullan99}).

Proto-solar and interstellar deuterium abudances thus bear the imprint
of BBN as well as the subsequent chemical evolution. Up to a few years
ago, they were the only available measurements of \dsh\ used to constrain 
BBN in a direct way. The situation has changed recently, 
as measurements of \dsh\ likely to be close to \dshprim\ 
have become possible (\eg,~Burles \& Tytler~\cite{bt1} \&~\cite{bt2}; 
Webb et al.~\cite{webb97}; see also 
Burles \& Tytler~\cite{burles_revue_98} for a~review). Taken
altogether, the abundances of deuterium seem to decrease with time, as
expected, although the dispersion remains rather large. These various
measurements and their trends are reviewed in detail in Lemoine et
al.~(\cite{lemoine99}), and we refer the reader to this review 
for more details.

The first measurements of the interstellar deuterium abundance \dshism, 
representative of the present epoch, were reported by Rogerson \& 
York~(\cite{ry73}) through Lyman absorption on the line of sight of 
$\beta$~Cen, using {\it Copernicus}. Their value of 
\dshism$\simeq1.4\pm0.2\times10^{-5}$ has not changed ever since 
and twenty years later, Linsky et al.~(\cite{linsky93} \&~\cite{linsky95}) 
measured \dshism~$=1.60 \pm 0.09 ^{+0.05}_{-0.10}\times 10^{-5}$ 
in the direction of Capella using HST-GHRS.
But it turns out that determinations of the \dshism\
ratio do not generally agree on a single value, even in the very local
medium (Vidal-Madjar et al.~\cite{avm78} \& \cite{avm86}, 
Murthy et al.~\cite{murty87} \&~\cite{murty90}). While many measurements 
are in agreement with the value of 
Linsky et al.~(\cite{linsky93} \&~\cite{linsky95}), many sightlines 
exhibit different values. For instance, \dshism$<10^{-5}$ toward 
$\lambda$~Sco (York~\cite{york83}), 
\dshism$\simeq7.\times10^{-6}$ toward $\delta$~Ori and 
$\epsilon$~Ori (Laurent et al.~\cite{laurent79}),  
\dshism$\simeq5.\times10^{-6}$ toward $\theta$~Car 
(Allen et al.~\cite{allen92}). 
Finally, thanks to new HST-GHRS observations 
of G191-B2B, Vidal-Madjar et al.~(\cite{avm98}) detected variations of 
\dshism\ by at least $\sim30$\% 
within the local interstellar medium on the sightline of G191-B2B, 
and Jenkins et al.~(\cite{jenkins99}) up to $\sim50$\% 
toward $\delta$~Ori, using IMAPS. We can also report the measurement 
of the 92cm hyperfine transition of \ion{D}{i} by Chengalur et 
al.~(\cite{chengalur97}), showing \dshism$\simeq3.9\pm1.0\times10^{-5}$. 
Although several scenarios have been proposed to explain these 
putative variations (\eg, Vidal-Madjar et al.~\cite{avm78}; 
Jura~\cite{jura}), the above measurements are still unaccounted 
for (Lemoine et al.~\cite{lemoine99}).

We introduced in Cycle~1 of HST a new type of target, white dwarfs in
the high temperature range, for which the depth of the \lya\ photospheric
absorption line is reduced, and whose stellar continuum remains
smooth.  These targets also allow the study of lines of other species, 
such as \ion{N}{i} and \ion{O}{i}, which are shown to be reliable
tracers of \ion{H}{i} in the ISM (Ferlet~\cite{ferlet81}; York et
al.~\cite{york_etal83}).  These targets may also be chosen close to the
Sun so that the \ion{H}{i} column density is not too high and the
velocity strucure of the line of sight not too complex. We have already
observed the white dwarf G191-B2B in HST Cycle~1 (Lemoine et
al.~\cite{lemoine96}) and Cycle~5 (Vidal-Madjar et al.~\cite{avm98}) .
Continuing that program, we present here new ultraviolet observations of 
Sirius~A and its white dwarf companion Sirius~B performed with HST-GHRS. 
Previous HST-GHRS observations of Sirius~A reveal the velocity structure 
including two components on this sightline 
(Lallement et al.~\cite{lalle}), and also the possible detection of a 
diffuse interstellar cloud boundary (Bertin et al.~\cite{bert1}) and the 
detection of a stellar wind from Sirius~A (Bertin et al.~\cite{bert2}). 

From a subset of these new observations, the identifications of the
Sirius~A emission lines have been made beforehand 
(van~Noort et al.~\cite{noort}). 
Here, following the presentation of these spectroscopic 
observations and the data reduction in 
Section~\ref{Observations_and_data_reduction}, we study the structure
of the line of sight in Section~\ref{Study_of_sightline_structure}, the
\lya\ aborption features including \dsh\ study in
Section~\ref{Study_of_the_lya_lines_toward_Sirius_A_and_Sirius_B}, and
the ionization and metal abundances in
Section~\ref{Analysis_of_the_metallic_lines}.  We finally discuss our
results in Section~\ref{Discussion}.

\section{Observations and data reduction}
\label{Observations_and_data_reduction}

\subsection{Observations}

Our observations of the stars Sirius~A and Sirius~B were performed 
with GHRS (the {\it Goddard High Resolution Spectrograph}) 
onboard the {\it Hubble Space Telescope} in November 1996, in the frame of 
Cycle~6 Guest Observer proposals ID~6800 and ID~6828. 
A first attempt was made in September 1996
but failed because the stars were not correctly located within the GHRS
entrance slit. The observations were finally repeated with a different
pointing strategy which was fully successful and allowed for extremely
good observations of both stars. The spectra were acquired at high and 
medium spectral resolution (Echelle-A and G140M gratings). 
Most of them have a very good quality and show clearly interstellar lines. 
The wavelengths ranges of the spectra are listed in Table~\ref{obs}.

\begin{center}
\begin{table*}
\caption[]{List of our GHRS spectra.}
\label{obs}
\begin{tabular}{lcccccccc}
\hline
\hline
Target & Spectral range & Elements & GHRS Grating & 
Exposition time & Date of observation & Proposal & \# \\
\hline
Sirius~A & 1188 \AA \ - 1218~\AA & N$\,\textsc i$, Si$\,\textsc {ii}$, 
Si$\,\textsc {iii}$, H$\,\textsc i$, D$\,\textsc i$ & 
G140M  & 1632.0 s & 1996 Nov. 20 & ID 6800 & 1 \\
Sirius~A & 1278~\AA \ - 1307~\AA & O$\,\textsc i$, Si$\,\textsc {ii}$  
& G140M  & 217.6 s  & 1996 Nov. 20 & ID 6800 & 2 \\
Sirius~A & 1308~\AA \ - 1337~\AA & C$\,\textsc {ii}$   & G140M        
& 217.6 s  & 1996 Nov. 20 & ID 6800 & 3 \\
Sirius~A & 1196~\AA \ - 1203~\AA & N$\,\textsc i$  & Echelle-A  
& 1305.6 s & 1996 Nov. 20 & ID 6800 & 4 \\	
Sirius~A & 1201~\AA \ - 1208~\AA & Si$\,\textsc {iii}$ & Echelle-A  
& 870.4 s  & 1996 Nov. 20 & ID 6800 & 5 \\	
Sirius~A & 1298~\AA \ - 1306~\AA & O$\,\textsc i$, Si$\,\textsc {ii}$ 
& Echelle-A  & 130.2 s  & 1996 Nov. 20 & ID 6800 & 6 \\	
Sirius~A & 1331~\AA \ - 1339~\AA & C$\,\textsc {ii}$ & Echelle-A  
& 217.6 s  & 1996 Nov. 20 & ID 6800 & 7 \\	
\hline
Sirius~B & 1188~\AA \ - 1218~\AA & N$\,\textsc i$, Si$\,\textsc {ii}$, 
Si$\,\textsc {iii}$, H$\,\textsc i$, D$\,\textsc i$ & 
G140M  & 1532.8 s  & 1996 Nov. 18 & ID 6828 & 8 \\
Sirius~B & 1278~\AA \ - 1307~\AA & O$\,\textsc i$, Si$\,\textsc {ii}$ 
& G140M  & 217.6 s  & 1996 Nov. 18 & ID 6828 & 9\\
Sirius~B & 1308~\AA \ - 1337~\AA & C$\,\textsc {ii}$ & G140M        
& 217.6 s  & 1996 Nov. 18 & ID 6828 & 10 \\
Sirius~B & 1196~\AA \ - 1203~\AA & N$\,\textsc i$ & Echelle-A  
& 3481.6 s & 1996 Nov. 21 & ID 6800 & 11 \\	
Sirius~B & 1201~\AA \ - 1208~\AA & Si$\,\textsc {iii}$ & Echelle-A  
& 1740.8 s & 1996 Nov. 18 & ID 6828 & 12 \\
Sirius~B & 1212~\AA \ - 1219~\AA & H$\,\textsc i$, D$\,\textsc i$ 
& Echelle-A  & 2067.2 s & 1996 Nov. 18 & ID 6828 & 13 \\	
Sirius~B & 1212~\AA \ - 1219~\AA & H$\,\textsc i$, D$\,\textsc i$ 
& Echelle-A  & 3481.6 s & 1996 Nov. 20 & ID 6800 & 14 \\	
Sirius~B & 1298~\AA \ - 1306~\AA & O$\,\textsc i$, Si$\,\textsc {ii}$ 
& Echelle-A  & 217.6 s  & 1996 Nov. 18 & ID 6828 & 15 \\	
Sirius~B & 1331~\AA \ - 1339~\AA & C$\,\textsc {ii}$ & Echelle-A  
& 217.6 s  & 1996 Nov. 18 & ID 6828 & 16 \\	
\hline
\hline
\end{tabular}
\end{table*}
\end{center}

The G140M grating provides a resolving power 
$R=\lambda/\Delta\lambda\simeq20,000$, \ie\ a spectral resolution 
of $\sim15$~\kms. With the Echelle-A grating we have obtained a resolving 
power $R\simeq85,000$, \ie\ a spectral resolution of $\sim3.5$~\kms .
We used only the Small Science Aperture (SSA), corresponding to
0.25" on the sky and illuminating one diode to achieve the best possible
resolving power. For further details on the instrumentation, see 
Duncan~\cite{duncan}).

\subsection{Data reduction}
\label{Data_reduction}

Our data were reduced with the Image Reduction and Analysis Facility 
(IRAF) software, using the STSDAS package.

During the observations, we used the FP-SPLIT mode which splits the total
exposure time into successive cycles of four sub-exposures, each 
corresponding to a slightly different projection of the spectrum on the 
photocathode. We used the ``quarter stepping'' mode, which provides a 
sample of 4 pixels per resolution element. This allows simultaneously
to oversample the spectrum, since, for instance, the SSA does not fulfill 
the Nyquist sampling criterion, and to correct for the granularity of 
the photocathode. The effect of the photocathode on each diode being 
the same for the four sub-exposures, it is possible to evaluate this 
granularity from the comparison of the four sub-exposures where a 
constant granularity effect mixes with a non-constant photon 
statistical noise. 

\begin{figure}
\resizebox{\hsize}{!}{\includegraphics{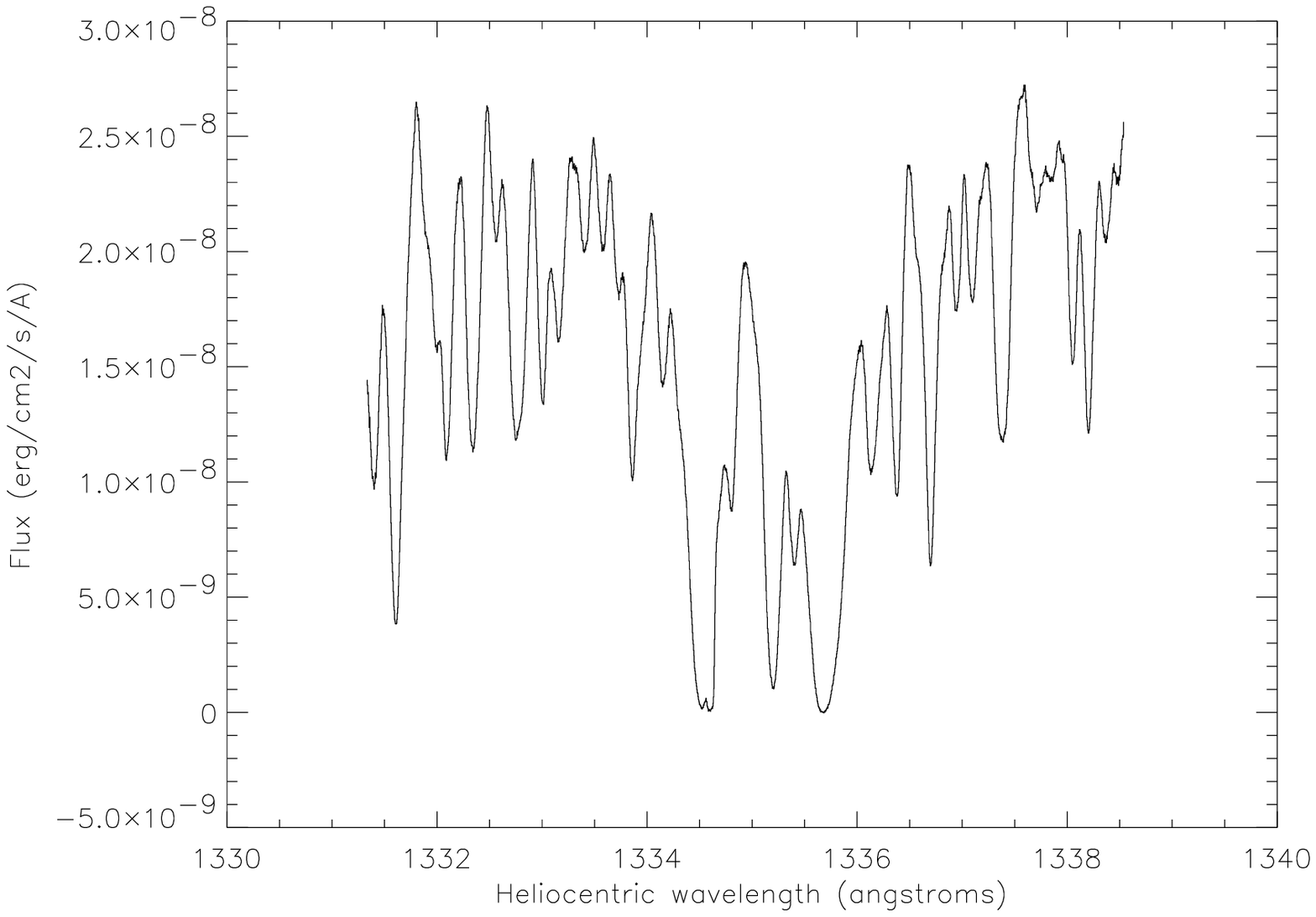}}
\resizebox{\hsize}{!}{\includegraphics{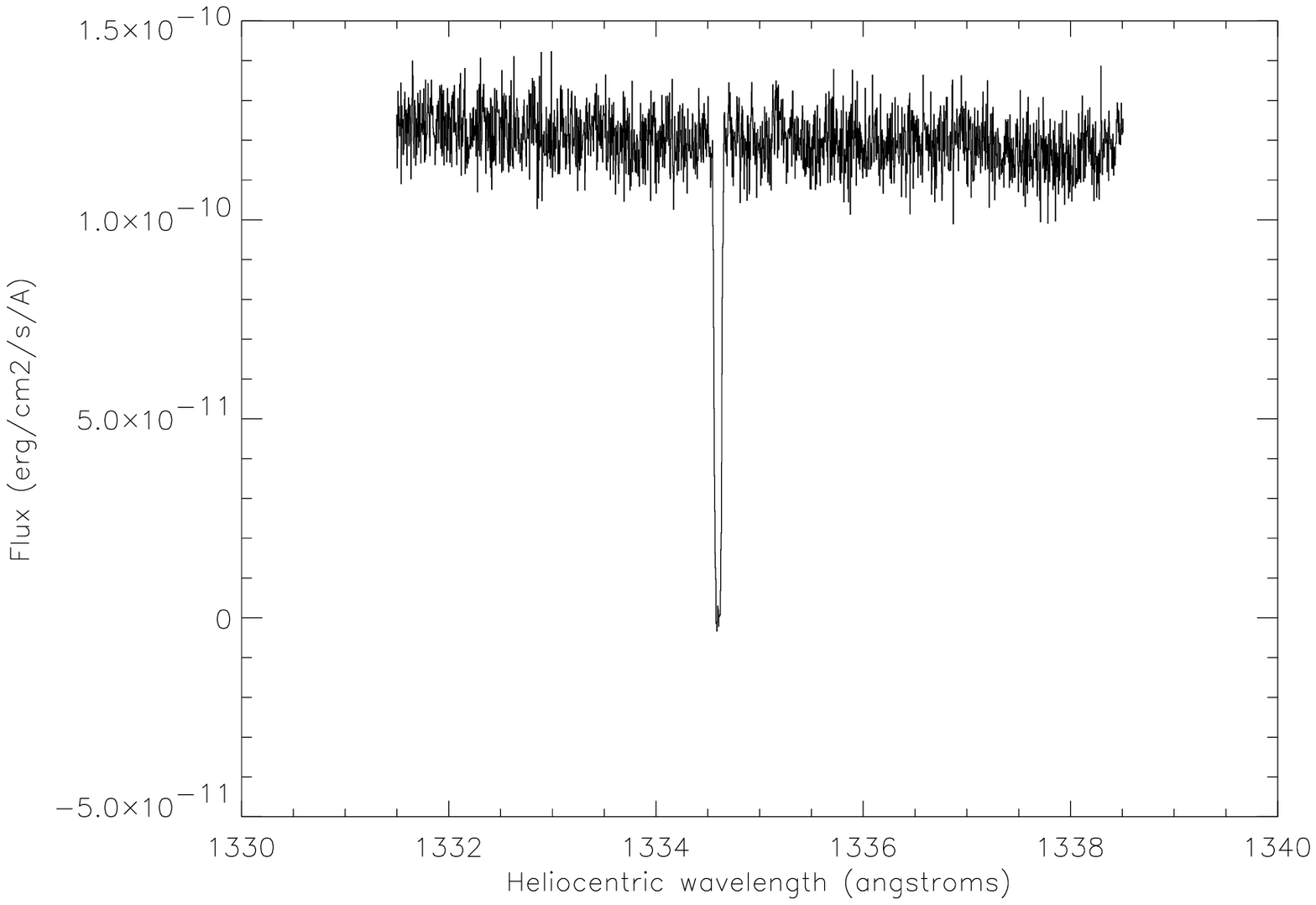}}	
\caption[]{Examples of spectra obtained with HST-GHRS Echelle-A: 
\ion{C}{ii} area toward Sirius~A (top, spectrum \#~7) and toward Sirius~B 
(bottom, spectrum \#~16). 
In the Sirius~A spectrum we see a lot of stellar lines, and an 
interstellar absorption on the red wing of the \ion{C}{ii}
1334.5~\AA\ stellar line. In the Sirius~B spectrum we see only the
interstellar \ion{C}{ii} 1334.5~\AA\ line on a very flat stellar 
continuum.}
\label{Examples1}
\end{figure}

\begin{figure}
\resizebox{\hsize}{!}{\includegraphics{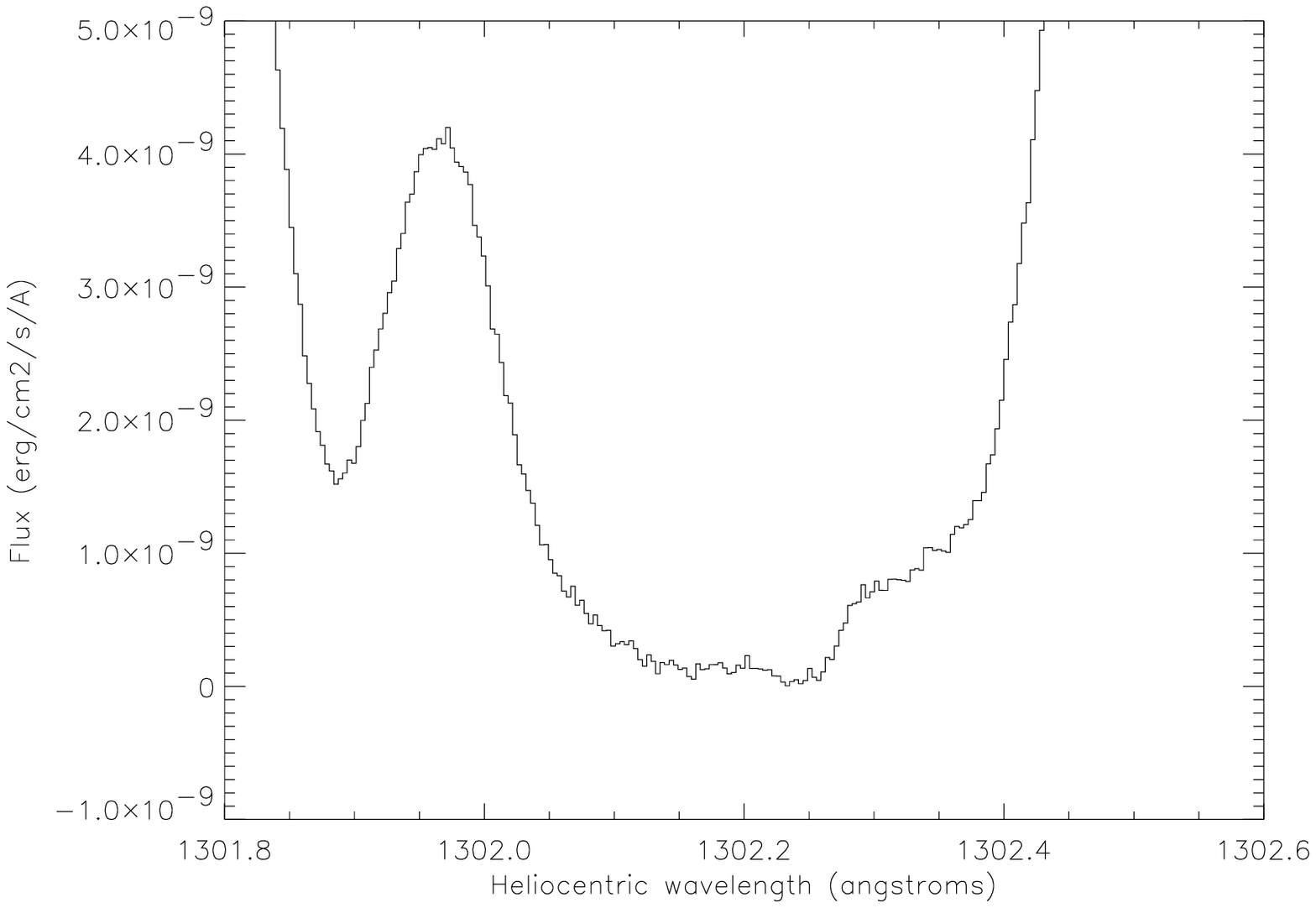}}
\resizebox{\hsize}{!}{\includegraphics{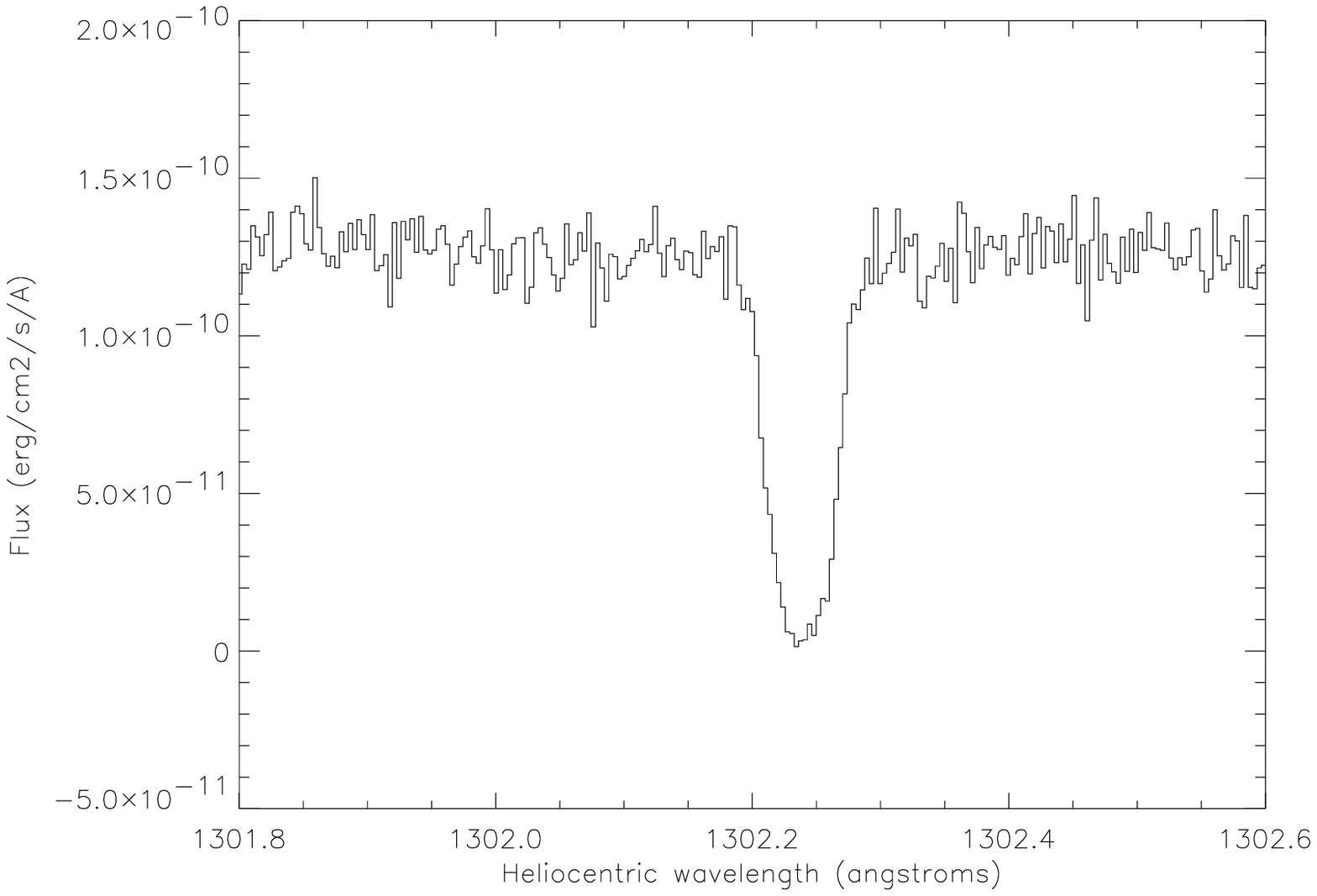}}
\caption[]{Examples of spectra obtained with HST-GHRS Echelle-A
(magnification): \ion{O}{i} area toward Sirius~A (top, spectrum \#~6) and 
toward Sirius~B (bottom, spectrum \#~15). On the Sirius~A spectrum, the
interstellar absorption line (the small feature at 1302.25~\AA) is at the 
bottom of a stellar line. For this small feature, the continuum is low 
and less simple than in the case of Sirius~B spectrum, on which we see
only the interstellar line, on a very flat stellar continuum.}
\label{Examples2}
\end{figure}

We use several methods to average the four series of sub-exposures 
to obtain the final spectra. The main point was to focus the alignment
between the different sub-exposures, in order to avoid 
any artificial broadening of the lines, or even erasing the weaker ones.
Where the signal-to-noise ratio ($S/N$) was high enough, 
we used the standard method to average the sub-exposures and 
correct for the granularity, which is available in the IDL GHRS package,
under the {\it corre\_hrs} procedure. This procedure, which executes an
automatic auto-correlation between the sub-exposures, is efficient when
the sub-exposures are bright enough.
We found that this simple treatment allowed us to reach good alignment for 
ten of our spectra (spectra \#~1, 2, 3, 4, 6, 7, 8, 9, 10 and 15 in 
Table~\ref{obs}). In the remaining cases the $S/N$ was too low, and the 
{\it corre\_hrs} procedure was not able to identify and then correlate
features on the sub-exposures. For these six low $S/N$ spectra 
(spectra \#~5, 11, 12, 13, 14 and 16 in Table~\ref{obs}), we chose to
estimate the shifts between the different sub-exposures one by one, after
smoothing them in order to increase the $S/N$, and then to average the 
sub-exposures with the right shifts (and with no additional 
smoothing). The shifts found thus were very close to the nominal ones 
in the FP-SPLIT mode. However these 
six low $S/N$ spectra are not corrected for the photocathode granularity. 
Instead we have searched for such possible defects by adding the 
sub-exposures corresponding a priori to the same spectral instrument 
shift, improving then the $S/N$, and building in such a way 
four different shifted spectra where the photocathode
defects appear at fixed positions. We found no defects near the observed
spectral lines. This can be seen, for example, in the Fig.~\ref{Examples1} 
and Fig.~\ref{Examples2} on which are shown 
the Sirius~A and Sirius~B Echelle-A data corresponding 
to both \ion{C}{ii} 1334~\AA \ and \ion{O}{i} 1302~\AA\ areas.

In our high and medium spectral resolution data, we have detected 10 
interstellar lines toward Sirius~A and/or Sirius~B: \ion{N}{i} 1200~\AA\ 
triplet, \ion{O}{i} 1302~\AA, \ion{C}{ii} 1334~\AA, \ion{Si}{ii} 
1190~\AA, 1193~\AA\ and 1304~\AA, \ion{D}{i} 1215~\AA \ and \ion{H}{i} 
\lya\ (see Table~\ref{detect}).

To properly deduce interstellar abundances from interstellar
absorption lines, it is extremely important to precisely evaluate the
zero flux level. The zero flux level may be slightly erroneous due to 
the scattered light, which is greater in the case of an \'echelle grating 
like Echelle-A than in the case of a classical grating like G140M, 
because of the diffuse light produced from the adjacent spectral orders. 
Thus, under the two assumptions that the zero flux is well known for 
G140M spectra, and that the width of the line spread function
with Echelle-A is negligible in comparison with the one obtained with  
G140M, we used the G140M spectra to adjust the zero level of 
the corresponding Echelle-A spectra (for exemple, we adjust spectrum 
\#~15 with \#~9, spectrum \#~16 with \#~10...). Accordingly, after
degrading the Echelle-A spectrum to the one of G140M 
\ie\ projection of Echelle-A spectrum on pixels with size equal to 
those of G140M spectrum, and then convolution with the G140M 
line spread function), we achieve the flux shift between the degraded 
Echelle-A spectrum and the corresponding G140M one. 
We corrected thus the zero flux levels of the non-degraded Echelle-A 
spectra with these shifts, in the cases of unsaturated lines. 
The most obvious cases are the broad and saturated \lya\ lines
for which the central saturated cores should be at zero flux level. 
We shift these spectra in order to adjust the \lya\ central cores at 
zero flux level (the level of that core has been 
found to be negative after the standard pipeline correction). 
This obvious case allows us to test and confirm that the zero 
flux level is well known in G140M spectra. It permits to ensure 
that the precision on the zero flux level for the G140M spectra, 
and thus for the corrected Echelle-A ones, is better than 3\% 
of the stellar continuum.

Seven of the Echelle-A exposures (\#~4, 5, 12, 13, 14, 15 and 16) 
were immediately preceded by a platinum lamp calibration exposure 
and thus allowed us to calculate the residual wavelength shift left 
by the standard {\it calib\_hrs} calibration procedure. Together with 
the oversampling mode, this correction allowed us to reach an absolute 
calibration accuracy of $\pm 1.5$~\kms\ in radial velocity (the radial
velocities will thereafter be given in heliocentric frame). 

We chose not to rebin our spectra \ie\ they present four pixels per
resolution element approximately, which is the standard sampling mode. 
For our fits (\S~\ref{Study_of_sightline_structure} and 
\S~\ref{Study_of_the_lya_lines_toward_Sirius_A_and_Sirius_B}) we then 
used a gaussian PSF, with a FWHM slightly greater than four pixels.

In order to study the interstellar contributions, the regions near the
interstellar spectral lines were normalized by the stellar continuum to
unity. This could be done without difficulty, except in the case of the 
\lya\ line toward Sirius~B, with polynomial whose degree and parameters 
were chosen using the procedure described in 
Lemoine et al.~(\cite{lemoine}). The photospheric \lya\ continua of 
Sirius~B was non-trivial and its normalization will be described in 
\S~\ref{Study_of_the_lya_lines_toward_Sirius_A_and_Sirius_B}.

The \lya\ high resolution spectra of Sirius~A obtained using 
Echelle-A was unusable, probably because of the very high diffusion of 
adjacent order in the spectrograph by this bright star. 
The \lya\ high resolution spectra of Sirius~B obtained using 
Echelle-A presented a very low $S/N$ ($\sim5$). We did not use them 
in our fits of the \lya\ lines 
(\S~\ref{Study_of_the_lya_lines_toward_Sirius_A_and_Sirius_B}), which are 
made only on the medium spectal resolution data obtained using G140M. We 
just use them in order to check that the geocoronal \lya\ emission during 
our observations was at the bottom of the saturated \lya\ absorptions. 
Indeed, this emission is detected in the Sirius~B \lya\ Echelle-A 
spectra, but not in the G140M spectra. The geocoronal emission thus 
did not deteriorate the wings of the \lya\ absorption lines.

\section{Structure of the line of sight}
\label{Study_of_sightline_structure}

As it can be seen in 
\S~\ref{Study_of_the_lya_lines_toward_Sirius_A_and_Sirius_B}, 
the \lya\ lines are not obvious. We thus used only the interstellar 
metal lines (\ie\ all but \lya ) to study the structure of the
line of sight toward the binary Sirius~A~/~Sirius~B. We assume in all 
our study that the interstellar absorbers are the same toward Sirius~A 
and Sirius~B, the two stars being separated by less than 4~arcsec on 
the sky at the time of our observations, which corresponds to 
$\sim10$~AU at the Sirius distance of 2.6~pc. 

For a given transition of a given element, each component in the 
sightline produces an absorption line modeled by a Voigt profile, 
which is defined, in addition to atomic parameters, by four cloud 
parameters: the radial velocity $v$ (in \kms) of the cloud, the column 
density $N_e$ (in cm$^{-2}$) of element $e$, the temperature $T$ 
(in K) of the gas, and its turbulent velocity $\sigma$ (in~\kms). 

In order to determine the number of clouds, their velocity,
temperature, turbulence and columns density for each element, we used a
new fitting program developed by M.~Lemoine, which permits to fit Voigt
profiles to several lines in different spectral ranges simultaneously.
This new program works in the same spirit as the code presented in
Lemoine et al.~(\cite{lemoine}), which obtains the best fit by $\chi^2$
simulated annealing optimization. This allows us to find the best
solution compatible with different spectral ranges based on the basic
assumption that all the considered lines give the same values for $v$,
$T$ and $\sigma$ for a given component, and all the lines for a given
element $e$ and a given component give the same value for $N_e$. The
spread of the lines combines both parameters $T$ and $\sigma$, which
can only be separately determined  if several elements with different
masses are simultaneously fitted. This  procedure was used to solve the
structure of the line of sight toward the white dwarf G191-B2B
(Vidal-Madjar et al.~\cite{avm98}).

\begin{center}
\begin{table*}
\caption[]{Interstellar spectral lines detected on HST-GHRS spectra 
toward Sirius~A and Sirius~B. $S/N$ is the signal to noise ratio per 
pixel at the continuum of the interstellar lines. 
The atomic data are from Morton~(\cite{morton}).}
\label{detect}
\begin{tabular}{lccccccc}
\hline
\hline
Element & Wavelength (\AA) & Oscillator & Spontaneous transition 
& Target & Spectral & $S/N$ & Reference \\
& (vacuum) & strength & probability (s$^{-1})$ & & resolution \\
\hline
\hline
N$\,\textsc i$ & $1199.5496$ & $1.328\times10^{-1}$ & $4.104\times10^8$ &
Sirius~A & $85,000$ & $5$ & This work \\
 & & & & Sirius~B & $20,000$ & $40$ & This work \\
 & & & & Sirius~B & $85,000$ & $20$ & This work \\
\hline
N$\,\textsc i$ & $1200.2233$ & $8.849\times10^{-2}$ & $4.097\times10^8$ &
Sirius~A & $85,000$ & $5$ & This work \\
 & & & & Sirius~B & $20,000$ & $45$ & This work \\
 & & & & Sirius~B & $85,000$ & $20$ & This work \\
\hline
N$\,\textsc i$ & $1200.7098$ & $4.423\times10^{-2}$ & $4.093\times10^8$ &
Sirius~A & $85,000$ & $5$ & This work \\
 & & & & Sirius~B & $20,000$ & $35$ & This work \\
 & & & & Sirius~B & $85,000$ & $15$ & This work \\
\hline
O$\,\textsc i$ & $1302.1685$ & $4.887\times10^{-2}$ & $3.204\times10^8$ &
Sirius~A & $85,000$ & $10$ & This work \\
 & & & & Sirius~B & $20,000$ & $50$ & This work \\
 & & & & Sirius~B & $85,000$ & $15$ & This work \\
\hline
C$\,\textsc {ii}$ & $1334.5323$ & $1.278\times10^{-1}$ 
& $2.393\times10^8$ & Sirius~A & $85,000$ & $50$ & This work \\
 & & & & Sirius~B & $20,000$ & $40$ & This work \\
 & & & & Sirius~B & $85,000$ & $20$ & This work \\
\hline
Si$\,\textsc {ii}$ & $1190.4158$ & $2.502\times10^{-1}$ 
& $5.888\times10^8$ & Sirius~B & $20,000$ & $50$ & This work \\
\hline
Si$\,\textsc {ii}$ & $1193.2897$ & $4.991\times10^{-1}$ 
& $2.338\times10^9$ & Sirius~B & $20,000$ & $45$ & This work \\
\hline
Si$\,\textsc {ii}$ & $1304.3702$ & $1.473\times10^{-1}$ 
& $5.776\times10^8$ & Sirius~A & $85,000$ & $10$ & This work \\
 & & & & Sirius~B & $20,000$ & $45$ & This work \\
 & & & & Sirius~B & $85,000$ & $15$ & This work \\
\hline
Fe$\,\textsc {ii}$ & $2344.2141$ & $1.097\times10^{-1}$ 
& $1.664\times10^8$ &
Sirius~A & $85,000$ & $200$ & Lallement et al.~(\cite{lalle}) \\
\hline
Fe$\,\textsc {ii}$ & $2600.1729$ & $2.239\times10^{-1}$ 
& $2.209\times10^8$ &
Sirius~A & $85,000$ & $300$ & Lallement et al.~(\cite{lalle}) \\
\hline
Mg$\,\textsc {ii}$ & $2796.3521$ & $6.123\times10^{-1}$ 
& $2.612\times10^8$ &
Sirius~A & $85,000$ & $150$ & Lallement et al.~(\cite{lalle}) \\
\hline
Mg$\,\textsc {ii}$ & $2803.5310$ & $3.054\times10^{-1}$ 
& $2.592\times10^8$ &
Sirius~A & $85,000$ & $200$ & Lallement et al.~(\cite{lalle}) \\
\hline
D$\,\textsc i$ & $1215.3394$ & $4.165\times10^{-1}$ & $6.270\times10^8$ &
Sirius~A & $20,000$ & $20$ & This work \\
 & & & & Sirius~A & $15,000$ & $20$ & Lallement et al.~(\cite{lalle}) \\
 & & & & Sirius~B & $20,000$ & $10$ & This work \\
 & & & & Sirius~B & $85,000$ & $5$ & This work \\
\hline
H$\,\textsc i$ & $1215.6701$ & $4.164\times10^{-1}$ & $6.265\times10^8$ &
Sirius~A & $20,000$ & $20$ & This work \\
 & & & & Sirius~A & $15,000$& $20$ & Lallement et al.~(\cite{lalle})  \\
 & & & & Sirius~B & $20,000$ & $10$ & This work \\
 & & & & Sirius~B & $85,000$ & $5$ & This work \\
\hline
\hline
\end{tabular}
\end{table*}
\end{center}

The previous study of the line of sight toward Sirius~A 
(Lallement et al.~\cite{lalle}) clearly showed two distinct interstellar 
clouds with a projected velocity shift equal to $5.7\pm0.2$~\kms. 
The red component was identified as the Local Interstellar Cloud (LIC), 
detected in many directions with projected velocities corresponding to 
the coherent motion of a cloud in which the solar system is embedded 
(Lallement \& Bertin~\cite{lall_bert_lic}).
To improve the precision of our study, we decided to fit the 10 lines of 
our new observation together with the \ion{Fe}{ii} 2344~\AA, \ion{Fe}{ii} 
2600~\AA, \ion{Mg}{ii} 2796~\AA\ and \ion{Mg}{ii} 2803~\AA\ lines
from Lallement et al.~(\cite{lalle}), 
which present very high $S/N$ (see Table~\ref{detect}).

\subsection{Velocity shifts correction}
\label{Velocity_shifts_correction}

\begin{figure} 
\resizebox{\hsize}{!}{\includegraphics{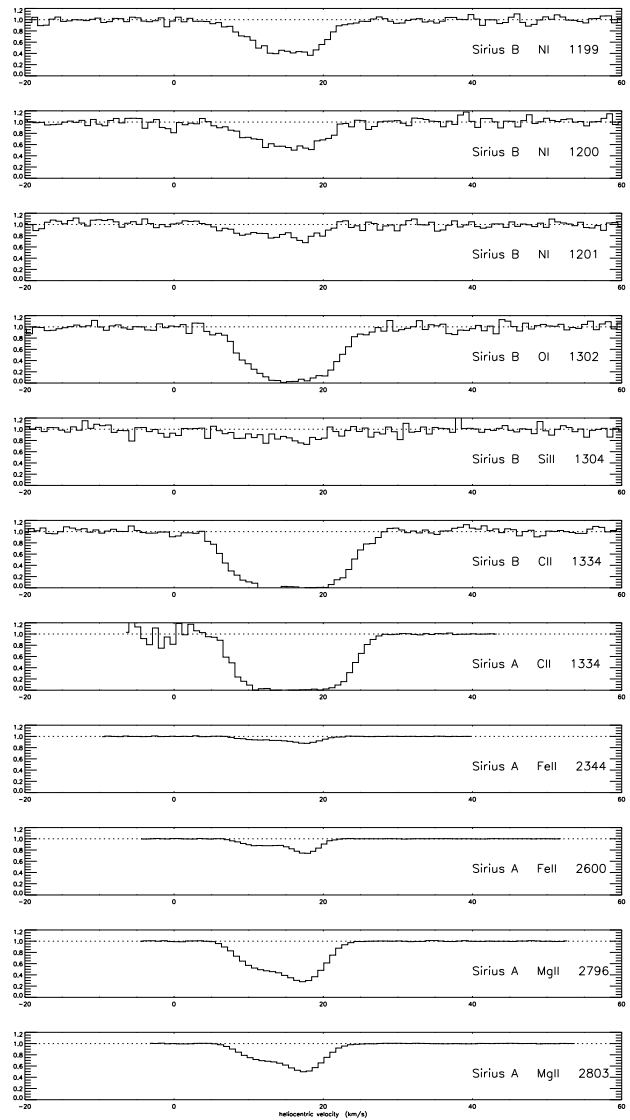}}  
\caption[]{High resolution interstellar lines toward Sirius~A and 
Sirius~B after velocity shifts correction (solid lines), and 
normalized stellar continuum (dotted lines).}
\label{traceHR}
\end{figure}

\begin{figure}     
\resizebox{\hsize}{!}{\includegraphics{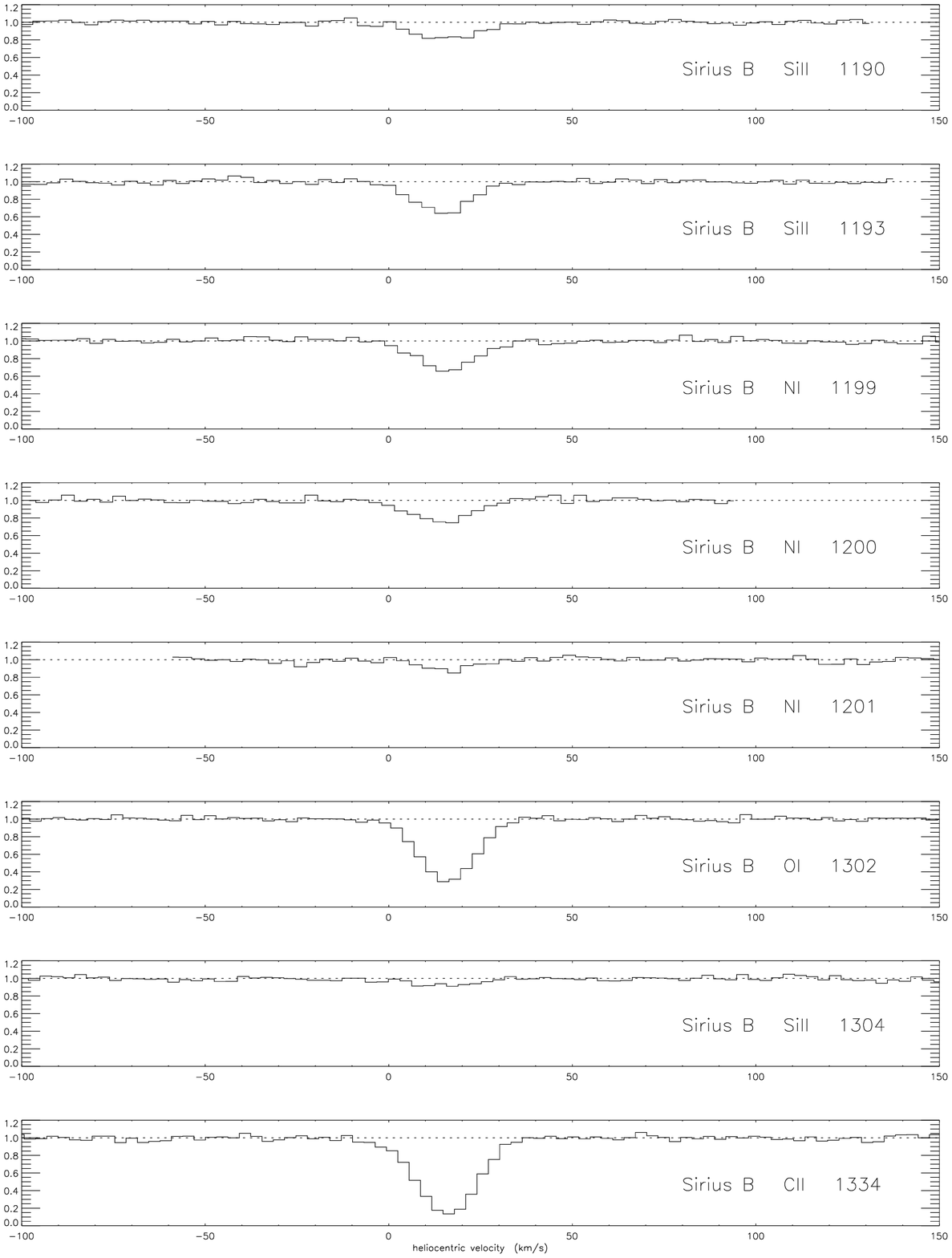}}  
\caption[]{Medium resolution interstellar lines toward Sirius~B 
after velocity shifts correction (solid lines), and normalized stellar 
continuum (dotted lines).}
\label{traceMR}
\end{figure}

Our fitting program is suitable for a simultaneous study of several
lines with the restriction that their wavelengths are precisely
determined. If there exist significant velocity shifts between
different lines caused for instance, by instrumental effects, the
program then becomes unable to find the right solution coherent with
all the fitted lines. Therefore, on a first iteration, we have corrected
for any possible instrumental velocity shifts.

In order to estimate the values of these instrumental shifts, 
we decided to begin our study by fitting one by one all the lines, 
assuming that there were two components, the blue one (BC) and the 
red one (LIC), with the following constraints:

\begin{itemize}
\item
$\Delta v_{\mathrm{LIC-BC}}=5.7$~\kms,
\item   
$1.5$~\kms $\leq \sigma_{\mathrm{BC}}\leq 3.5$~\kms,
\item  
$100$~K$\leq T_{\mathrm{BC}}\leq 10000$~K,
\item   
$0.1$~\kms $\leq \sigma_{\mathrm{LIC}}\leq 2.5$~\kms,
\item  
$T_{\mathrm{LIC}}=7000$~K.
\end{itemize}

These values were the result of the Lallement et al.~(\cite{lalle}) 
paper. The strong constraints were the projected velocity shift 
$\Delta v_{\mathrm{LIC-BC}}=5.7$~\kms exactly, which is the 
Lallement et al.~(\cite{lalle}) result without the reported errors, 
and $T_{\mathrm{LIC}}=7000$~K, which is the standard value assumed 
for the LIC [see for example Linsky et al.~(\cite{linsky95})].

Each fit gives a pair of velocies, one for the BC and one for the LIC, 
separated by 5.7~\kms.

We did not find the same pairs of velocities for all the lines, and we 
interpreted these shifts as signs of instrumental errors in the absolute 
wavelength calibration. Indeed, the shift between the \ion{C}{ii} and 
\ion{O}{i} lines, for which the spectra was preceded by a platinum lamp 
calibration exposure, was on the order of 1~\kms, rather than for the 
lines not preceded by calibration exposures which presented greater 
shifts. We took as reference the average pair of velocities found 
with the spectra preceded by calibration exposures, and we shifted 
all our spectra by hand on this reference. 

We thus fit together the lines of higher $S/N$ and spectral 
resolution, and we obtained a first set of results for the values 
of $v$, $T$, $\sigma$ and $N_e$ for the two components. Fitting one 
by one the lower $S/N$ or lower spectral resolution data with the 
contrainsts from these first results, we improved the velocity 
shift correction in the same spirit as above.

After this iterative process, there remained no instrumental 
velocity shift greater than $\pm 0.5$~\kms\ in our data, 
as it can be seen in Fig.~\ref{traceHR} (high resolution 
data) and in Fig.~\ref{traceMR} (medium resolution data). 
All the detected interstellar lines were thus well 
calibrated in wavelength, and ready to be fitted all together 
to constraint the sightline.

\subsection{Interstellar structure of the line of sight}
\label{Interstellar_structure_of_the_line_of_sight}

We fitted together 19 high and medium spectral resolutions interstellar 
lines but this time without the hypotheses of  
\S~\ref{Velocity_shifts_correction} on the radial velocity shift 
between the two clouds or on the widths of the lines. Assuming only 
that there are two components on the sightline, we let all the 
other parameters vary. We did not include in this fit the high spectral 
resolution interstellar lines \ion{N}{i}, \ion{O}{i} and \ion{Si}{ii} 
toward Sirius~A due to their low $S/N$. 

\begin{figure*}
\resizebox{\hsize}{!}{\includegraphics{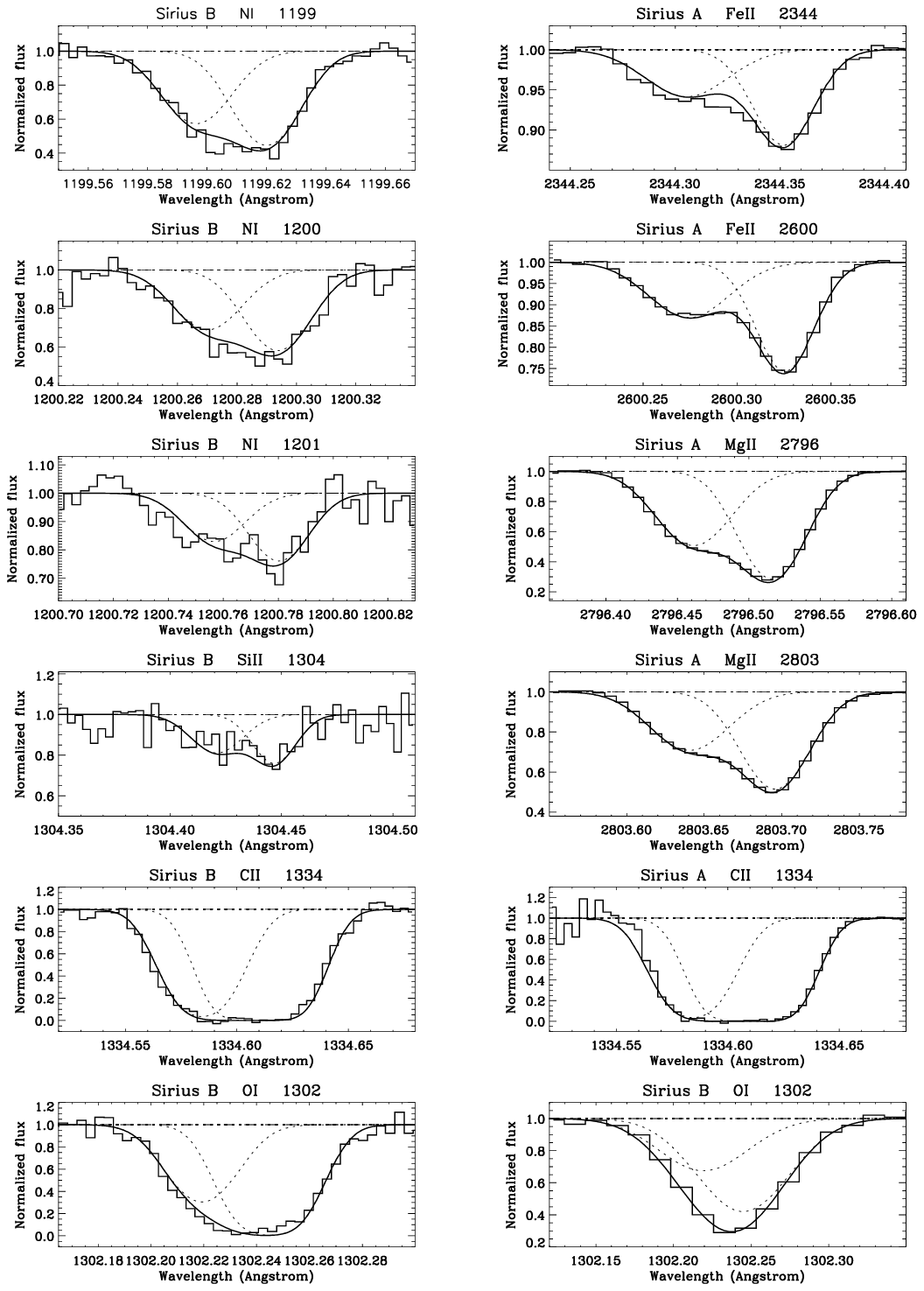}}   
\caption[]{The final fit for the metal lines toward Sirius~B and 
Sirius~A (see values found in Table~\ref{res_19raies}).
All the high spectral resolution lines are plotted, together with one 
medium spectral resolution line (Sirius~B \ion{O}{i}, bottom right).}
\label{tous_les_fits}
\end{figure*}

The values found from this global fit are reported in  
Table~\ref{res_19raies} and the fit plots in Fig.~\ref{tous_les_fits}.
The reduced \ki2\ from this fit is 1.12 for 749 degrees of freedom. 
The error bars given in Table~\ref{res_19raies} are statistical 
$\pm2\sigma$ errors according to the $\Delta$\ki2\ method. We checked 
them on the one hand by trial and error, and on the other hand 
by performing fits to all species but one, each species 
being excluded in turn, and comparing the obtained results.

Adopting the same method/procedure, but using a fitting program developed 
by D.~Welty (Welty et al.~\cite{welty}), we get very good agreement in 
the column densities, temperatures and velocity separation of BC and LIC. 

The quality of the fit allows to affirm that there is no signature of an 
extra component on this line of sight. Thus we assume in the 
following that there are only two interstellar clouds toward Sirius.

\begin{center}
\begin{table}
\caption[]{Results of the final fit of the 19 interstellar metal 
lines together (see plots in Fig.~\ref{tous_les_fits}).}
\label{res_19raies}
\begin{tabular}{c|cc}
\hline
\hline
\\
Component & BC & LIC \\
\\
\hline
\\
Temperature [K] & $3000^{+2000}_{-1000}$ & $8000^{+500}_{-1000}$ \\
\\
Turb. velocity [\kms] & $2.7\pm0.3$ & $0.5\pm0.3$ \\
\\
Radial velocity & $11.7\pm1.5$ & $17.6\pm1.5$ \\
 \ [\kms] & 
\multicolumn{2}{|c}{$\Delta v_{\mathrm{LIC-BC}}=5.9\pm0.3$} \\
\\
$N($\ion{N}{i}$)$ [cm$^{-2}$]   & $9.2^{+2.0}_{-1.0}\times 10^{12}$
 & $1.3^{+0.1}_{-0.2}\times 10^{13}$ \\
\\
$N($\ion{O}{i}$)$ [cm$^{-2}$]   & $5.0^{+1.5}_{-1.0}\times 10^{13}$
 & $3.4^{+1.5}_{-1.0}\times 10^{14}$ \\
\\
$N($\ion{Si}{ii}$)$ [cm$^{-2}$] & $2.7^{+1.0}_{-0.5}\times 10^{12}$
 & $3.0^{+1.0}_{-0.5}\times 10^{12}$ \\
\\
$N($\ion{C}{ii}$)$ [cm$^{-2}$]  & $6.0^{+2.5}_{-1.5}\times 10^{13}$
 & $4.2^{+3.0}_{-2.0}\times 10^{14}$ \\
\\
$N($\ion{Fe}{ii}$)$ [cm$^{-2}$]  & $5.5^{+0.4}_{-0.4}\times 10^{11}$
 & $8.7^{+0.3}_{-0.3}\times 10^{11}$ \\
\\
$N($\ion{Mg}{ii}$)$ [cm$^{-2}$] & $1.0^{+0.1}_{-0.1}\times 10^{12}$
 & $1.7^{+0.1}_{-0.1}\times 10^{12}$ \\
\\
\hline
\hline
\end{tabular}
\end{table}
\end{center}

Our results are in agreement with those of 
Lallement et al.~(\cite{lalle}), 
which provided for temperatures and turbulent velocities  
$T_{\mathrm{LIC}}=7600\pm3000$~K and 
$\sigma_{\mathrm{LIC}}=1.4^{+0.6}_{-1.4}$~\kms, and 
$T_{\mathrm{BC}}=1000^{+6000}_{-1000}$~K and 
$\sigma_{\mathrm{BC}}=2.9^{+0.1}_{-0.5}$~\kms.
These two components were already also detected toward $\epsilon$~CMa 
which is located at 12\degr\ from Sirius 
(Gry et al.~\cite{gry95}), ``component~2'' in that study being the 
same as the one named here ``BC''. The values by 
Gry et al.~(\cite{gry95}) toward $\epsilon$~CMa are 
$T_{\mathrm{LIC}}=7200\pm2000$~K and 
$\sigma_{\mathrm{LIC}}=2.0\pm0.3$~\kms, and 
$T_{\mathrm{BC}}=3600\pm1500$~K and 
$\sigma_{\mathrm{BC}}=1.85\pm0.3$~\kms. 
The agreement is good for the temperatures, but less good for the 
turbulent velocities. Recent observations of the LIC given 
toward Capella 
$T_{\mathrm{LIC}}=7000\pm500\pm400$~K and 
$\sigma_{\mathrm{LIC}}=1.6\pm0.4\pm0.2$~\kms\ 
(Linsky et al.~\cite{linsky95}), 
toward Procyon 
$T_{\mathrm{LIC}}=6900\pm80\pm300$~K and 
$\sigma_{\mathrm{LIC}}=1.21\pm0.27$~\kms\ 
(Linsky et al.~\cite{linsky95}), and toward the white dwarf G191-B2B 
$T_{\mathrm{LIC}}=4000^{+2000}_{-1500}$~K and 
$\sigma_{\mathrm{LIC}}=2.0^{+0.5}_{-1.0}$~\kms\ 
(Vidal-Madjar et al.~\cite{avm98}). 
Here, again the agreement with our values is good for temperature, but 
less good for turbulent velocity. That might suggest a different turbulent 
structure toward Sirius.

\section{Analysis of the \lya\ lines toward Sirius~A and Sirius~B}
\label{Study_of_the_lya_lines_toward_Sirius_A_and_Sirius_B}

The problem of possible instrumental radial velocity shifts is the same  
for the \lya\ spectra here as  with the metal lines above. Being in 
the same spectrum as the triplet \ion{N}{i}, we corrected the Sirius~B 
G140M \lya\ line by the shift found for that triplet. It was less 
obvious for the Sirius~A G140M \lya\ line because the interstellar 
\ion{N}{i} lines were not detected on this spectra due to the too low 
spectral resolution and $S/N$. But in spite of low $S/N$, the interstellar 
\ion{N}{i} triplet was detected in the Sirius~A Echelle-A spectra, on 
the red wings of the stellar \ion{N}{i} lines. Thus it was possible to 
align the interstellar \ion{N}{i} high resolution lines in the sightline 
of Sirius~A with the equivalent high resolution lines in the sightline 
of Sirius~B, for which instrumental velocity shifts were already 
corrected (see above). With \ion{N}{i} triplet and \lya\ being on the 
same G140M spectrum, alignement of the Sirius~A G140M \ion{N}{i} triplet 
spectral zone with the Sirius~A Echelle-A \ion{N}{i} triplet spectral 
zone further permitted alignment of the Sirius~A G140M \lya\ line.

However, a check on the deuterium absorption lines in the blue 
wings of both \lya\ lines indicates a velocity shift between the two 
spectra. It is probably due to the large wavelength coverage in G140M 
spectra between the \ion{N}{i} and \lya\ lines and to the low $S/N$ 
on the high resolution Sirius~A \ion{N}{i} spectrum, implying a coarse 
determination of interstellar component velocities. Here, using again 
the same strategy to correct velocity shifts, we fit one by one the 
two deuterium lines, assuming the results obtained above for the 
structure of the sightline. We found a shift 1.5~\kms\ between the 
two \ion{N}{i} lines, a shift which was corrected by hand.

\begin{figure}      
\resizebox{\hsize}{!}{\includegraphics{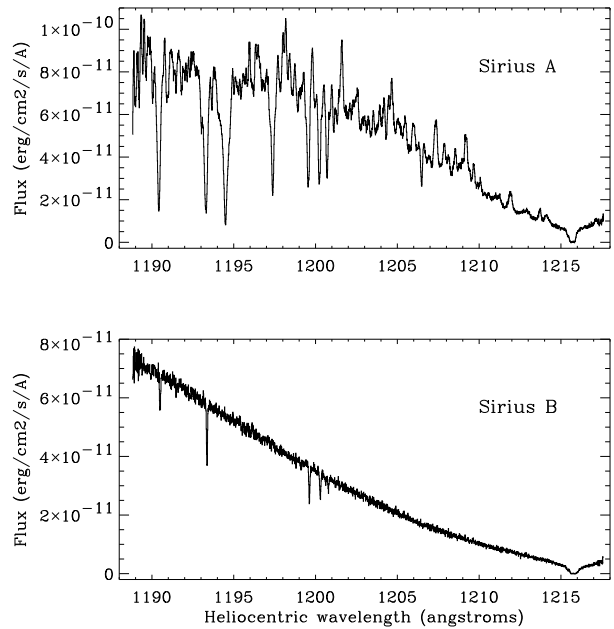}}  
\caption[]{G140M spectra of the \lya\ region of Sirius~A (top) and 
Sirius~B (bottom). 
Emission lines in the blue wing of the Sirius~A \lya\ profile was  
studied by van~Noort et al.~(\cite{noort}). Superimposed on the 
photospheric \lya\ absorption lines of the both stars are the lines 
\ion{Si}{II} (1190~\AA\ and 1193~\AA ) and \ion{N}{i} 
(triplet at 1200~\AA ). Whereas the origin of these five lines is 
interstellar toward Sirius~B, they are photospheric toward Sirius~A, 
the interstellar components being not resolved.
Only the interstellar \lya\ line is resolved both toward Sirius~A 
and Sirius~B on these spectra, at the bottom of the photospheric 
\lya\ lines, near 1216~\AA\ on both plots.}
\label{2_lyman}
\end{figure}

After all shift corrections, we estimate that the possible instrumental 
velocity shift between the Sirius~A and Sirius~B G140M \lya\ lines was 
lower than $\pm 5$~\kms. The two corrected spectra in this range from 
Sirius~A and Sirius~B are plotted on the Fig.~\ref{2_lyman}. 
Interstellar absorption lines are detected at the bottom of the 
photospheric \lya\ of both stars.

\begin{figure}      
\resizebox{\hsize}{!}{\includegraphics{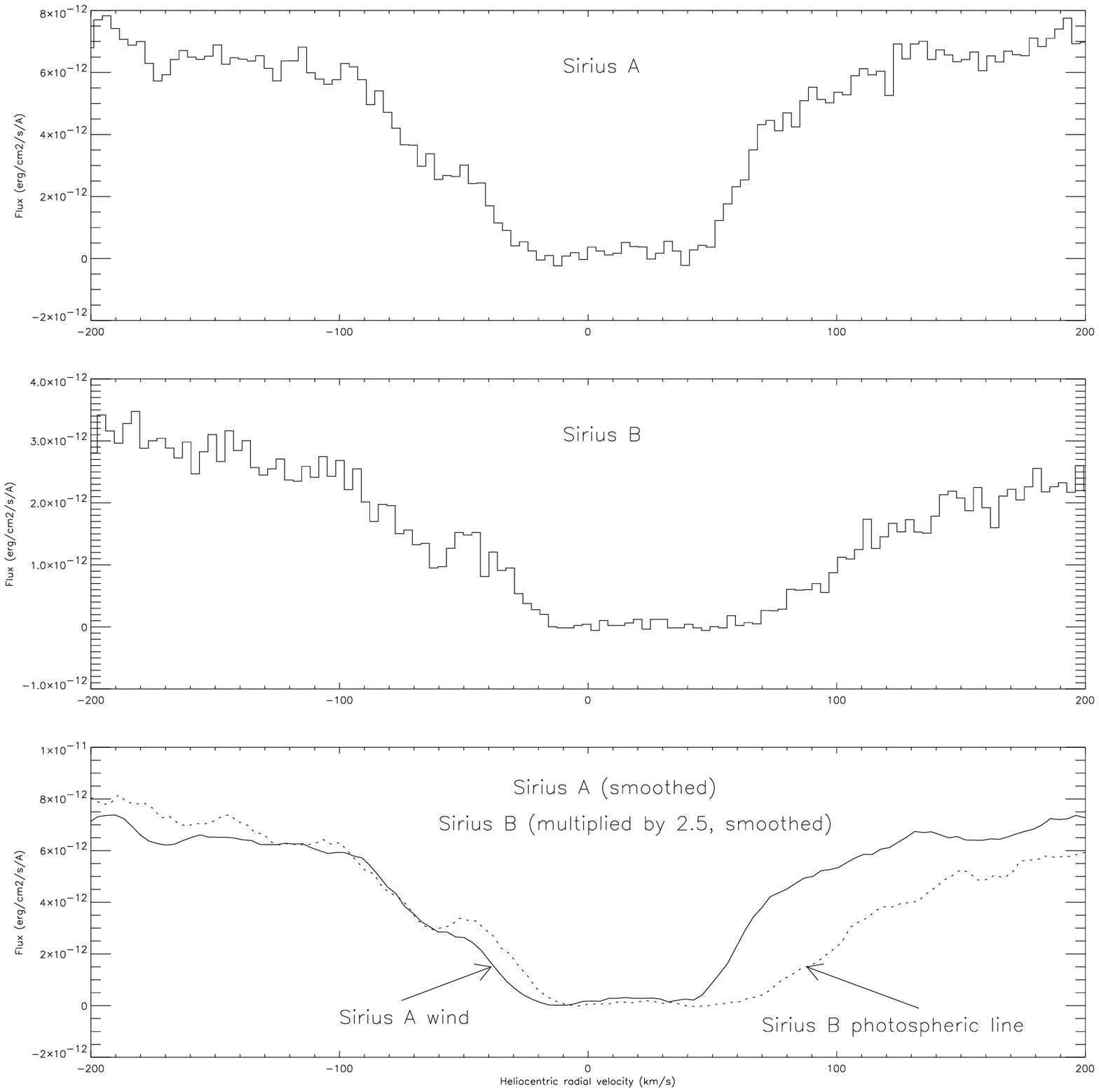}}  
\caption[]{Comparison of the \lya\ interstellar absorption lines toward 
Sirius~A and Sirius~B. 

{\it Top plot}: Sirius~A \lya\ line with G140M. Deuterium absorption is 
visible on the blue wing of the \lya\ line, near $-60$~\kms.

{\it Middle plot}: Sirius~B \lya\ line with G140M. Compared to the 
Sirius~A one, the deuterium absorption is more contrasted, and the red 
wing of the \lya\ line is more extended.

{\it Bottom plot}: Sirius~A (solid line) and Sirius~B (dotted line) 
smoothed spectra (overplot of the both above spectra). The flux of 
the Sirius~B spectra is here multiplied by 2.5 to overlap the 
Sirius~A one. Instrumental velocity shift between these two 
spectra are smaller than $\pm5$~\kms. That is confirmed by the 
good superposition of the two deuterium lines on this plot. 
The absorption excess in the blue wing of the Sirius~A \lya\ 
line is interpreted as signature of stellar wind from Sirius~A, 
and the absorption excess in the red wing of the Sirius~B \lya\ 
line as signature of the white dwarf photospheric absorption 
\lya\ line.}
\label{3_lyman_alpha}
\end{figure}

As one can see on the magnification of these two lines (see
Fig.~\ref{3_lyman_alpha}), however the profiles of the interstellar
\lya\ lines are not the same toward Sirius~A and Sirius~B.  The
deuterium feature is detected on the blue wing toward both stars, but
it is more contrasted on the Sirius~B spectrum. The interstellar
\lya\ line toward Sirius~A seems to be more extended in the blue wing
by at least 10~\kms.  In the same way, the red wing of the Sirius~B
interstellar \lya\ is more extended by at least 50~\kms\ compared to
the one toward Sirius~A.

The differences between the two \lya\ interstellar absorption profiles
are not caused by a simple instrumental wavelength shift. Indeed, the
sizes of the extend differences are greater than the precision on the
relative velocity ($\pm 5$~\kms), which is confirmed by the relatively
good superposition of the two deuterium lines on the blue wings of
\lya.  Moreover, beyond the extent difference, the shape difference of
the two profiles allows us to affirm that it is not a simple shift.

Comparing the high spectral resolution spectra of the \ion{C}{ii} 
1334~\AA\ and \ion{O}{i} 1302~\AA\ interstellar aborption lines, we did 
not detect such profile difference between Sirius~A and Sirius~B as in 
the case for the 1200~\AA\ \ion{N}{i} triplet lines. The \ion{N}{i}
triplet observations toward Sirius~A, however, have a low $S/N$.

We thus observed an absorption excess in the blue wing of the
\lya\ interstellar line toward Sirius~A, and an absorption excess in
the red wing of the \lya\ interstellar line toward Sirius~B. These two
excesses are only detected in the \lya\ line.

The two stars being separated by less than 4~arcsec on the sky at the
time of our observations, which corresponds to $\sim10$~AU at the
Sirius distance of 2.6~pc, the processes which cause these excesses are
likely due to the stars themselves, or to circumstellar material very
close to the stars. Processes having origins in interstellar or
interplanetary media, or even in Earth environment, are very unlikely
in order to explain differences on such small scales.

We interpret these signatures in the \lya\ profiles as 
stellar wind from Sirius~A and of the core of the Sirius~B \lya\ 
photospheric absorption line respectively, 
both superimposed on the classic interstellar 
feature. In \S~\ref{The_Sirius_B_photospheric_lya_line} and 
\S~\ref{Confirmation_of_the_stellar_wind_from_Sirius_A} 
we discuss further these interpretations.

\subsection{The Sirius~B photospheric \lya\ line}
\label{The_Sirius_B_photospheric_lya_line}

Photospheric lines from Sirius~B must be centered near 
70~\kms\ (heliocentric). Indeed, the gravitational redshift, 
measured by Greenstein et al.~(\cite{greenstein}) at 
$89\pm 16$~\kms\ (note that the value obtained theoretically
using the mass and radius is about 81~\kms), have to be corrected by 
the proper and orbital velocities of the white dwarf, which was about 
$-20$~\kms\ for the sum of both motions at the epoch of observation 
[orbital data of Sirius~B  by Couteau~\& Morel can be found in 
Benest~\& Duvent~(\cite{benest})]. As seen in Fig.~\ref{3_lyman_alpha}, 
the velocity position of the red absorption excess of the Sirius~B 
\lya\ line is of this order of magnitude and may explain the 
excess by the Doppler core of the \lya\ photospheric absorption from 
Sirius~B. This explanation requires fewer assumptions
than any other, like infall on the white dwarf, for example.

\begin{figure}      
\resizebox{\hsize}{!}{\includegraphics{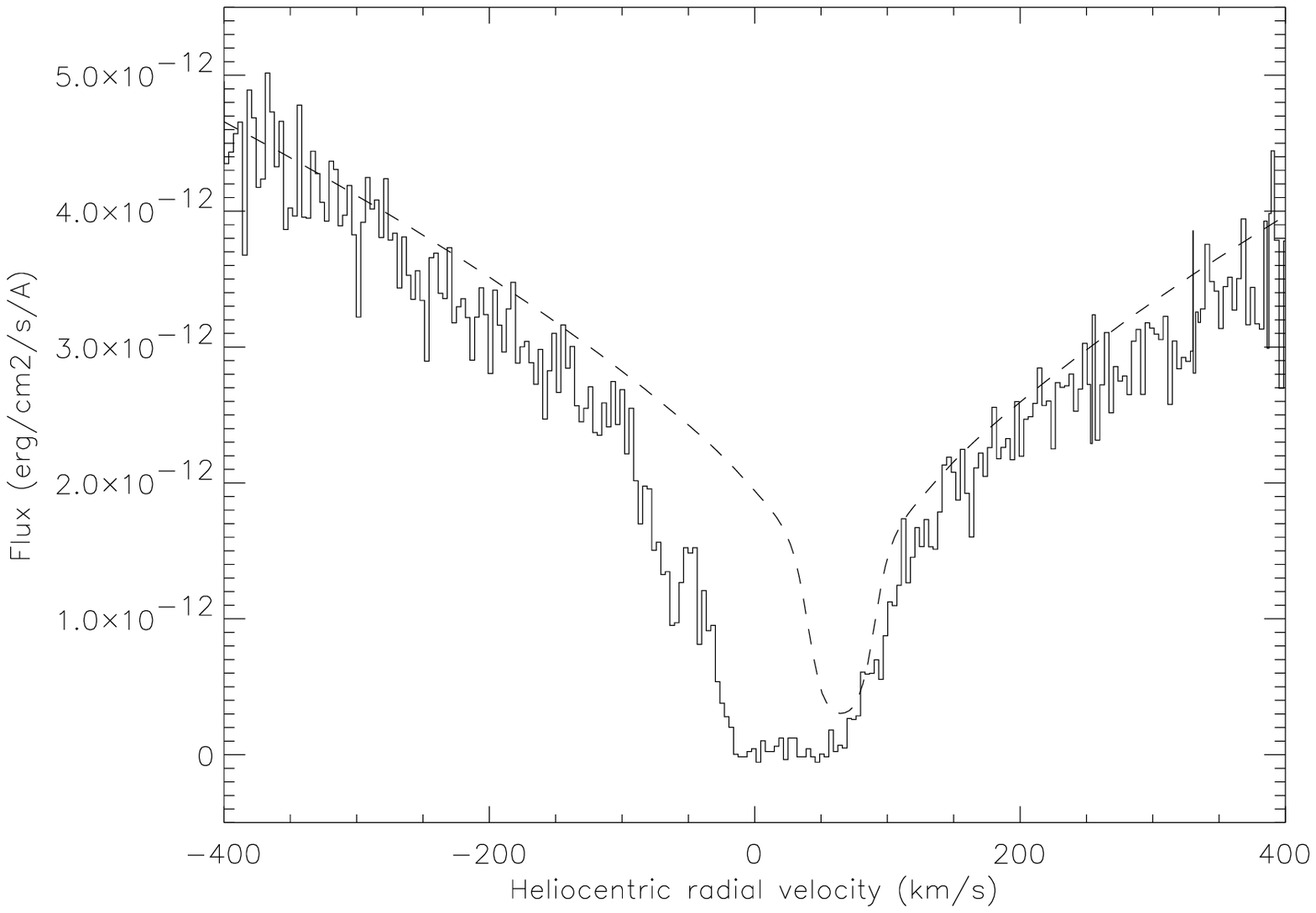}}   
\caption[]{\lya\ photosperic continuum for Sirius~B. The model 
(dashed line) is shifted by 65~\kms\ to fit the data (solid line).}
\label{fitdk}
\end{figure}

We fit the Sirius~B \lya\ photospheric profile by LTE and NLTE models 
calculated by the team of D.~Koester using the very accurate
parameters that Holberg et al.~(\cite{holberg98}) have derived
for Sirius~B: effective temperature $T_{\rm eff}=24790$~K and surface 
gravity log~$g=8.57$. As seen in Fig.~\ref{3_lyman_alpha}, the core of 
the \lya\ Sirius~B photospheric line falls very near the zero flux 
level, perhaps even reaches it. We thus choose the model which presents 
the deepest core, \ie\ the NLTE model. The best velocity shift found 
to fit the data is $v=65$~\kms, with an estimated uncertainty of 
$\pm15$~\kms. The plot of the fit is shown on Fig.~\ref{fitdk}.

The observed Sirius~B \lya\ spectrum normalized by this model must thus 
show  only the interstellar absorption due to both BC and LIC clouds 
identified in \S~\ref{Interstellar_structure_of_the_line_of_sight}. 
With the stellar continuum shape being uncertain, and the velocity 
shift (65~\kms , see Fig.~\ref{fitdk}) 
not determined better than $\pm15$~\kms, the red 
wing of this normalized interstellar line is very uncertain. 
However it is superimposed rather well on the red wing of the Sirius~A 
\lya\ line, as expected if our interpretation is right. 

The Sirius~B \lya\ blue wing on the other hand is well known, since the 
stellar continuum is relatively flat on this part. This is the blue wing 
we will use below for the ``pure interstellar'' spectrum
(\S~\ref{The_interstellar_D/H_ratio_in_the_line_of_sight_of_Sirius}).

\subsection{Confirmation of the stellar wind from Sirius~A}
\label{Confirmation_of_the_stellar_wind_from_Sirius_A}

Bertin et al.~(\cite{bert2}) have reported earlier the feature in the blue 
wing of the interstellar \lya\ line toward Sirius~A, and interpreted it as 
due to stellar wind. Our confirmation of this observation toward Sirius~A 
and the fact that we do not observe this feature toward Sirius~B is 
in agreement with this interpretation.

The comparison of the blue wings of the \lya\ lines of Sirius~A 
and Sirius~B could allow to estimate the shape of the absorption due to 
the wind, at least its blue part, its red part being lost in 
the saturated core of the two \lya\ lines. The wind absorption ranges at 
least from $-10$ to $-60$~\kms , in the range where 
Bertin et al.~(\cite{bert2}) detected it.

The wind absorption velocity distribution may be guessed assuming the 
velocity is in the rest frame of Sirius~A at its surface and increases 
its blue shift outwards until it reaches a velocity limit. However, 
the abundance of neutral relative to ionized hydrogen probably 
changes with distance from the star.

Thus fits of the Sirius~A wind absorption using a simple Voigt profile 
with a given velocity is too naive and could not give credible results. 
The study of the blue absorption feature is beyond the scope of the 
present work and will be the subject of a forthcoming paper.

\subsection{The interstellar \dsh\ ratio in the line of sight of Sirius}
\label{The_interstellar_D/H_ratio_in_the_line_of_sight_of_Sirius}

If both absorption excesses actually have a stellar origin, the
interstellar part of the \lya\ line toward Sirius is at most the
overlap between these two lines. Since each of the suspected
processes that affect one wing of the \lya\ line cannot affect the 
other wing, we assumed that the interstellar contribution to the \lya\ 
line on this line of sight, generated by the interstellar clouds BC 
and LIC found in \S~\ref{Interstellar_structure_of_the_line_of_sight}, 
is the overlap between these two lines (see below).

The Sirius~A \lya\ spectral region was normalized 
to unity by using a second degree polynomial stellar continuum 
[degree and parameters of that polynomial were found using the 
procedure described in Lemoine et al.~(\cite{lemoine})].
Using that Sirius~A \lya\ normalized spectrum and the Sirius~B \lya\ 
spectrum normalized by the photospheric model 
(see~\S~\ref{The_Sirius_B_photospheric_lya_line} and Fig.~\ref{fitdk}), 
we constructed a composit spectrum from the overlap of the two 
normalized spectra, \ie\ with the blue wing of the Sirius~B \lya\ line 
and the red wing of the Sirius~A one. If our interpretations of the 
two excesses are correct, and if there are no other components in the 
interstellar \lya\ line than the two indentified in the metal lines 
in \S~\ref{Interstellar_structure_of_the_line_of_sight}, this 
``pure interstellar'' spectrum presents absorption caused only by 
\ion{H}{i} and \ion{D}{i} from the BC and LIC and is free of any 
other processes.

We fit the ``pure interstellar'' spectrum in order to determine the
\ion{H}{i} and \ion{D}{i} column densities in BC and LIC, and thus the
D/H ratio in the two interstellar clouds. In these fits, we fixed the
values for the radial velocity shift, the temperatures and the
turbulent velocities of both interstellar clouds found using 
the 19 metal lines fit as described in
\S~\ref{Interstellar_structure_of_the_line_of_sight} 
(see Table~\ref{res_19raies}).

The first fit gave a \dshlic\ ratio in agreement with the 
Linsky et al.~(\cite{linsky95}) value of 
\dshlic~$=1.60 \pm 0.09 ^{+0.05}_{-0.10} \times 10^{-5}$, obtained toward 
Capella and Procyon, but a surprising \dshbc\ ratio about one order of 
magnitude lower, caused by a very low deuterium column density in the BC. 
We thus adopted in a first attempt the conservative method of 
adding an extra constraint by fixing the standard value 
\dshlic~$=1.6\times 10^{-5}$ in our \lya\ fits, and to let \dshbc\ free 
in order to study this ratio. We found then \dshbc$<0.5\times10^{-5}$.

The critical point which caused this very low \dsh\ ratio is the
\ion{D}{i} column density in the BC, which should not easily exceed 
$1 \times 10^{12}$~cm$^{-2}$ in the frame of our assumptions and 
as indicated by the \ki2. A \dsh\ ratio of 
order $1.6\times10^{-5}$ in BC thus requires a very low \ion{H}{i} 
column density in BC, implying a ratio $N_{\mathrm{LIC}}($\ion{H}{i}$) /
N_{\mathrm{BC}}($\ion{H}{i}$) \geq 10$.  This comes in contrast to a
ratio $N_{\mathrm{LIC}}($\ion{N}{i}$) / N_{\mathrm{BC}}($\ion{N}{i}$)
\simeq 1.4$, \ion{N}{i} column densities being accurate
because obtained from several non-saturated lines
(it is not the case for \ion{O}{i}). As \ion{N}{i} is
regarded as a rather good tracer of \ion{H}{i}
(Ferlet~\cite{ferlet81}), a ratio $N_{\mathrm{LIC}}($\ion{H}{i}$) /
N_{\mathrm{BC}}($\ion{H}{i}$)$ between 0.5 and 4 seems more realistic.

\ 

In order to obtain the most accurate results, we investigated and 
estimated the effects of possible systematic errors in our fits, 
possibly related to the set of assumptions made. 

\begin{figure}    
\resizebox{\hsize}{!}{\includegraphics{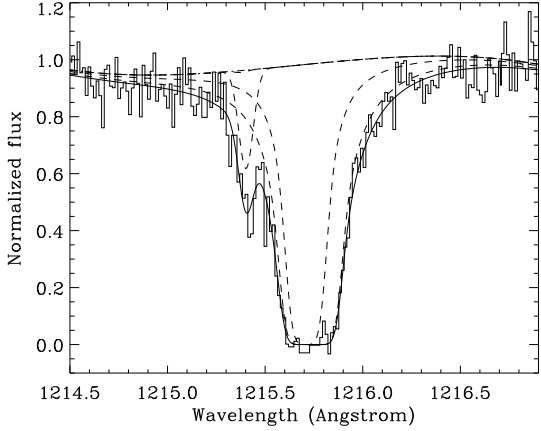}}  
\caption[]{Fit of the ``pure interstellar'' \lya\ line toward 
Sirius by a Voigt profil, the stellar continuum being fitted by 
a $3^{\rm rd}$~order polynomial. The continuum polynomial obtained 
is close to unity and interstellar column densities are close to those 
found if the stellar continuum is assumed to unity. 
Thus, systematic uncertainty in the stellar continuum does not seem to 
imply a serious uncertainty on our result.
The specrum fitted in Fig.~\ref{fit_id} and Fig.~\ref{pure_ism} 
is normalized by this $3^{\rm rd}$~order polynomial.}
\label{fit_polyn}
\end{figure}

The first cause of systematic uncertainty is the shape of the
\lya\ continuum.  Indeed, as discussed in
\S~\ref{The_Sirius_B_photospheric_lya_line}, the normalization of the
Sirius~B \lya\ line, whose center lies in the blue wing of the ``pure
interstellar'' \lya\ line, depends on both the choice of the stellar 
model and the velocity shift of the photospheric line, which is not 
known to better than $\pm 15$~\kms. Whereas the ``pure
interstellar'' spectrum was previously fitted without allowing 
the continuum level to vary during the fitting routine
(the continuum was fixed at unity), we then added more freedom by 
having a $3^{\rm rd}$ order polynomial in order to fit the
stellar continuum, this polynomial being fitted simultaneously 
with the Voigt
functions (see Fig.~\ref{fit_polyn}). 
Because the new corrected continuum is close to unity and \ion{H}{i} 
and \ion{D}{i} column densities are similar to the
ones found previously, this eliminates hidden systematics 
in the continuum as a serious compromise in our fit. In addition 
one should note that the inaccuracy of the fit related to the 
Sirius~B photospheric \lya\ line does not affect significantly 
the blue wing of the ``pure interstellar'' line, as already 
mentioned in \S~\ref{The_Sirius_B_photospheric_lya_line}.  
Although this correction was tiny, we normalized the ``pure 
interstellar'' spectra by this $3^{\rm rd}$ order polynomial 
to erase as well as possible the systmatics related to the 
profile, which may mimic interstellar damping wings.  

A second cause of systematic uncertainty is the instrumental 
velocity shift between the Sirius~A and Sirius~B G140M \lya\ lines, 
which can be $\pm 5$~\kms, as seen above. The shape of the 
``pure interstellar'' \lya\ line is compromised by this uncertainty. 
In order to quantify this effect we constructed two other 
``pure interstellar'' spectra, one in which both wings of the \lya\ 
were 5~\kms\ nearer in comparison with the nominal spectra, and one 
in which they were 5~\kms\ more distant. The fits of these two 
spectra did not provide very different 
results, and again produced a low \dshbc\ ratio.

We also estimated the influence of the zero flux level on the result.
By studying \ki2\ variations as a function of different zero flux levels, 
we found that the zero level of this saturated line is known to $\pm3\%$ 
of the continuum flux level, in agreement with our previous estimation 
(see \S~\ref{Data_reduction}).  Fitting lines with zero levels 3\% 
higher or 3\% 
lower gave again very similar result, with low \dshbc\ ratio.

The last systematic which we studied is the effect of the 
uncertainty on the structure of the sightline. Indeed, we fix 
in the fiting of the \lya\ line the values:
$T_{\mathrm{BC}}=3000$~K, $T_{\mathrm{LIC}}=8000$~K, 
$\sigma_{\mathrm{BC}}=2.7$~\kms, $\sigma_{\mathrm{LIC}}=0.5$~\kms, 
and $\Delta v_{\mathrm{LIC-BC}}=5.9$~\kms, found in 
\S~\ref{Interstellar_structure_of_the_line_of_sight} 
by the study of metal lines. We used the error bars on $T$, $\sigma$ 
and $\Delta v_{\mathrm{LIC-BC}}$ reported in Table~\ref{res_19raies} 
to perform extra fits with different constraints, \ie\ with components 
more or less broad, or more or less shifted from each other.
Once again, we found similar results and low \dshbc\ ratio.

\ 

\begin{figure}      
\resizebox{\hsize}{!}{\includegraphics{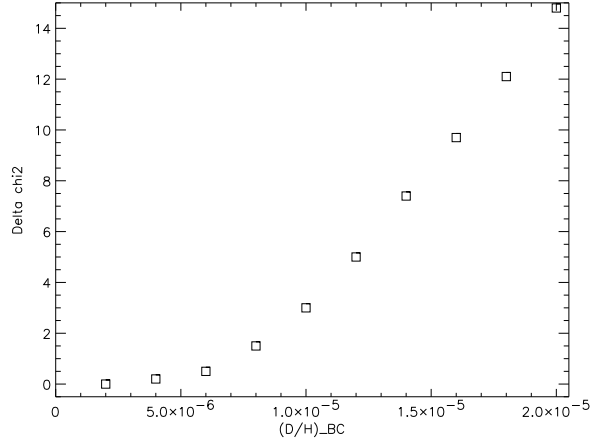}}   
\caption[]{\Deltaki2\ as a function of \dshbc\ [\dshlic\ being fixed at 
$=1.6\times10^{-5}$]. The fits are made on the ``pure interstellar'' 
spectrum. The criterion $0<$\Deltaki2$<10$ allow the range 
$0<$\dshbc$<1.6\times10^{-5}$.}
\label{detla_chi2}
\end{figure}

Whereas all these tests allowed us to estimate systematic error bars,
we evaluated statistical errors by the \Deltaki2\ method, fixing the
\dshbc\ to a value, and looking for the best \ki2\ for this given value,
and iterating. The evolution of \Deltaki2\ as a function of \dshbc\ is 
reported on Fig.~\ref{detla_chi2}. Following the same criterion as 
Vidal-Madjar et al.~(\cite{avm98}), we obtain \Deltaki2=10 for 
\dshbc$=1.6\times10^{-5}$. This statistical error is rather large, 
due to the relativelly low $S/N$ and spectral resolution. Possible 
systematic effects discussed above are thus negligible in comparison 
with this statistical uncertainty. However, we were able to apply this 
\Deltaki2\ method only to the ``pure interstellar'' spectrum, even 
though the blue wing of the Sirius~A \lya\ contains some information, 
and in particular also prohibits too high \dshbc\ values. 

We note that comparing many lines of sight through the LIC, 
Linsky~(\cite{linsky98}) recently estimates a mean value of
\dshlic$=1.50\times10^{-5} \pm0.10$. If we assume that value, 
we obtain \Deltaki2=10 for \dshbc$=1.7\times10^{-5}$.

\ 

\begin{figure}      
\resizebox{\hsize}{!}{\includegraphics{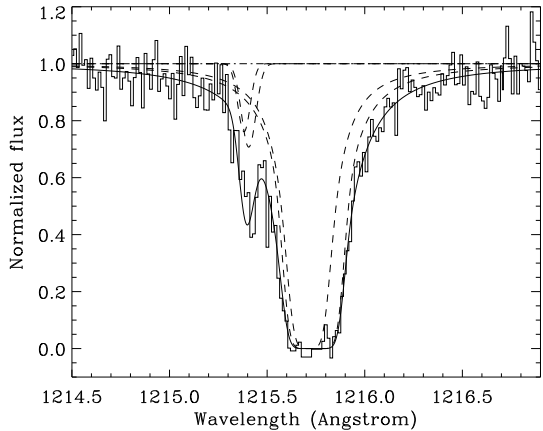}}   
\caption[]{Fit of the ``pure interstellar'' \lya\ line assuming 
\dshlic=\dshbc=$1.6\times10^{-5}$. The quality of the fit in the 
blue wing of the line is degraded in that case in comparison with 
the case assuming a lower \dshbc\ ratio (see Fig.~\ref{pure_ism}).}
\label{fit_id}
\end{figure}

We finally completed another study by fitting the ``pure interstellar'' 
\lya\ spectrum by relaxing the constraints on \dshlic,  
and assuming that the \dsh\ ratio is free but the same in the two 
components, in order to find which unique \dsh\ ratio 
is compatible with our data. While keeping a 
$N_{\mathrm{LIC}}($\ion{H}{i}$) / N_{\mathrm{BC}}($\ion{H}{i}$)$ 
close to unity as above, we found that unique ratio to be 
\dsh$=1.2\times10^{-5}$. Although we cannot exclude this result, 
this fit is significantly worse with a same \dsh\ ratio in both 
clouds, rather than with two different ones as above. Fits 
with higher common \dsh\ ratios ($1.4$ or $1.6\times10^{-5}$) degrade 
even more the quality of the fit (see Fig.~\ref{fit_id}) and could 
be rejected in term of \ki2. These data seems thus to reject the 
possibility of a unique \dshism\ ratio in both components although 
marginally. 

In brief, the results which we obtained in terms of \ion{H}{i} and 
\ion{D}{i} column densities, and \dsh\ ratios, for the BC and the 
LIC are summarized in Table~\ref{NH_ND_ISM}. The range which we 
finally obtain in the Blue Component for the deuterium abundance 
is $0<$\dshbc$<1.6\times10^{-5}$. The upper limit 
\dshbc$<1.6\times10^{-5}$ correponds to a \Deltaki2$=10$ in comparison 
with the lowest \ki2\ obtained for \dshbc$<0.5\times10^{-5}$. 
Thus we found that the deuterium abundance could be equal to 
$1.6\times10^{-5}$ in LIC and BC, but with a low probability according 
to our data. Moreover our \Deltaki2\ was obtained from the ``pure 
interstellar'' spectrum. Although we were unable to fit the deuterium 
Sirius~A line because of the stellar wind feature, that line shows  
also that the deuterium column density could not be too high. 
That suggests a lower upper limit for \dshbc . 

\begin{figure}      
\resizebox{\hsize}{!}{\includegraphics{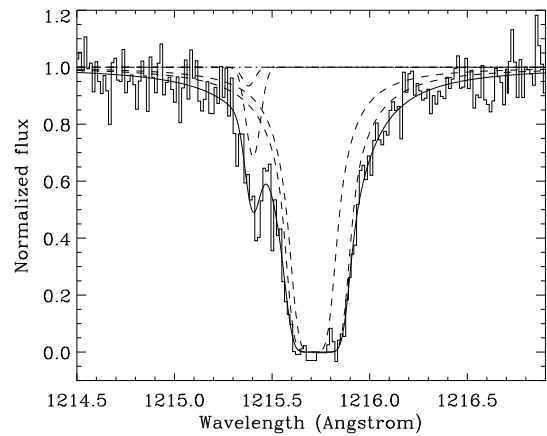}}  
\caption[]{Final fit of the ``pure interstellar'' \lya\ line toward 
Sirius, in which radial velocity shift $\Delta v_{\mathrm{LIC-BC}}$, 
temperatures and turbulent velocities of BC and LIC are fixed from 
Table~\ref{res_19raies}. \dshlic\ is fixed at $1.6 \times 10^{-5}$, 
and \dshbc\ at $0.5 \times 10^{-5}$.}
\label{pure_ism}
\end{figure}

Fig.~\ref{pure_ism} shows one of the best fits of the ``pure 
interstellar'' spectrum which we obtained, with the values 
\dshlic$=1.6\times10^{-5}$ and \dshbc$=0.5\times10^{-5}$.

\begin{center}
\begin{table}    
\caption[]{\ion{H}{i} and \ion{D}{i} column densities in BC and LIC 
clouds obtained by fitting of the ``pure interstellar'' \lya\ line
(see fit in Fig.~\ref{pure_ism}).}
\label{NH_ND_ISM}
\begin{tabular}{l|cc}
\hline
\hline
\\
Component & BC & LIC \\
\\
\hline
\\
$N($\ion{H}{i}$)$ [cm$^{-2}$]   &    $2.5^{+1.0}_{-1.0} \times 10^{17}$
                     &    $4.0^{+1.5}_{-1.0}  \times 10^{17}$ \\
\\
$N($\ion{D}{i}$)$ [cm$^{-2}$]   &    $1.0^{+2.0}_{-1.0} \times 10^{12}$ 
                     &    $6.5^{+1.5}_{-1.5}  \times 10^{12}$ \\
\\
\dsh                 &    $0.5^{+1.1}_{-0.5} \times 10^{-5}$
                     &    $1.6^{+0.4}_{-0.4} \times 10^{-5}$ \\
\\
\hline
\hline
\end{tabular}
\end{table}
\end{center}

\subsection{No-confirmation of the detection of a diffuse cloud boundary}

Bertin et al.~(\cite{bert1}) reported the detection of excess 
absorption in the red wing of the Sirius~A \lya\ line and 
interpreted it as being due to a warmer very diffuse neutral 
region, which may originate from an evaporative interface between 
the hot ``Local Bubble'' and the warm interstellar gas. 
Izmodenov et al.~(\cite{izmo}) proposed to explain that red 
excess in a different way, by the absorption of the neutralized, 
compressed solar wind from the heliosheath in the downwind direction.

We do not detect in our data the signature of that third absorber 
previously seen by Bertin et al~(\cite{bert1}) in the Sirius~A 
\lya\ line, despite very similar \lya\ profiles. 
This disagreement is neither due to 
a difference between the two spectra, caused for example by possible 
errors in the different data reduction procedures used for the 
different data sets gathered toward 
Sirius~A using the G160M grating (Bertin et al.~\cite{bert1}) 
and the G140M grating (our case), 
nor by real changes in the stellar line profile since both 
our \lya\ Sirius~A spectra are very similar. 
It is due to a different analysis of two almost identical 
spectra. In effect, the cause of this disagreement comes from 
the different \ion{D}{i} column densities 
we derive for the two components thanks to the Sirius~B \lya\ 
observation (clean over the \ion{D}{i} line), compared to the 
ones evaluated by Bertin et al.~(\cite{bert1}) from Sirius~A 
\lya\ alone. Indeed Bertin et al.~(\cite{bert1}) evaluated \ion{D}{i} 
column densities from Sirius~A first, and derived \ion{H}{i} column 
densities using the Linsky et al.~(\cite{linsky93}) \dshism\ ratio
as a unique common value. 
In addition to the fact that we do not find the same \dshism\ ratio 
for the two components, we argue that the \ion{D}{i} column 
densities obtained from Sirius~A \lya\ line are less reliable than 
those obtained from Sirius~B \lya\ line because of the blue excess 
observed toward Sirius~A. This excess, located near the wavelength 
of \ion{D}{i}, casts some doubts on the determination of the 
\ion{D}{i} column density by Bertin et al.~(\cite{bert1}).  

In effect 
we found a total \ion{D}{i} column density 1.4 times larger than the 
one found by Bertin et al.~(\cite{bert1}) and, 
according to our composite profile, a total \ion{H}{i} column 
density 1.9 times larger. This explains the differences  
between the two studies. Bertin et al.~(\cite{bert1}) also noticed 
that the \ion{H}{i} total column density they found was smaller than 
other evaluations in short lines of sight and in particular 
to the one obtained by Bruhweiler \& Kondo~(\cite{bruhweiler}) 
toward Sirius~B from the \ion{N}{i} 
column density using IUE: $N($\ion{H}{i}$)=8.5 \times 10^{17}$~cm$^{-2}$. 
Here we find a slightly lower total column density 
$N($\ion{H}{i}$)\sim6.5\times10^{17}$\cm2 .
Our value however agrees well with the recent result by 
Holberg et al.~(\cite{holberg98}) who found 
toward Sirius~B, from \ion{Si}{ii}, 
\ion{O}{i} and \ion{C}{ii} column densities: 
$N($\ion{H}{i}$)=5.2^{+1.4}_{-1.1} \times 10^{17}$ cm$^{-2}$.

In addition, we note that Izmodenov et al.~(\cite{izmo}) propose to 
explain the aborption excess in the blue wing of the \lya\ line 
toward Sirius~A by hydrogen atoms formed in an {\it astrosphere} 
around Sirius, similar to the {\it heliosphere} around the Sun. 
However, our data obtained toward Sirius~B show that 
we do not observe such a blue excess toward the Sirius~A companion. 
The two stars being separated by $\sim10$~AU at the time of our 
observations and the {\it siriosphere} having a size of about 200~AU, 
the {\it siriosphere} aborption should have been observed toward 
Sirius~A and Sirius~B with similar amounts of absorption. The fact 
that we observe it only toward Sirius~A favours the stellar wind 
interpretation rather than the {\it siriosphere} one. 
The Sirius~A wind proposed by Bertin et al.~(\cite{bert2}) is 
thus the more probable explanation since it was also detected 
in the \ion{Mg}{ii} doublet.

\section{Analysis of the metal lines}
\label{Analysis_of_the_metallic_lines}

In this section we discuss briefly the abundances measured in the 
metal lines.

\subsection{C, N, O and Si abundances and depletions}
\label{abundances_and_depletions}

Any discussion in the abundances and depletion level of C, N, O and Si 
is compromised by the uncertainties in the column density of \ion{H}{i} 
(especially of BC) and the metal lines (especially \ion{C}{ii}).

Nonetheless, comparing the column densities of the \ion{N}{i}, 
\ion{O}{i}, \ion{Si}{ii} and \ion{C}{ii} lines of the BC and the 
LIC relative to their respective \ion{H}{i} values, we determined 
the depletion factors as listed in Table~\ref{depletion}.

For BC, \ion{N}{i}, \ion{O}{i} and \ion{Si}{ii} 
show a depletion factor relative to the solar one 
(Anders \& Grevesse~\cite{ag89}) of $\sim2.5$ whereas 
\ion{C}{ii} is depleted by a smaller factor of $\sim1.2$. 
For LIC, \ion{N}{i} and \ion{Si}{ii} 
show a depletion factor of $\sim4.5$, \ion{C}{ii} 
is over abundant by a factor 2.5. In 
LIC, as opposed to BC, \ion{O}{i} does not show a similar depletion to 
\ion{N}{i} and \ion{Si}{ii}.

Instead of comparing column densities to \ion{H}{i} we used 
\ion{N}{i}, assuming \ion{N}{i} traces \ion{H}{i} well. We find 
depletions as shown in Table~\ref{depletion}. Again \ion{C}{ii} 
and \ion{O}{i} interstellar abundances appear too high in the LIC.

Linsky et al.~(\cite{linsky95}) mesured in the LIC toward Capella 
$10^6 \times N($\ion{O}{i}$)/N($\ion{H}{i}$)\simeq480$ from GHRS 
observations. Meyer et al.~(\cite{meyer_OI}) derived from 
GHRS observations the accurate average ISM gas neutral abundance 
$10^6 \times N($\ion{O}{i}$)/N($\ion{H}{i}$)=319\pm14$. Our values 
obtained in BC and LIC are respectively $200^{+100}_{-130}$ and 
$850^{+360}_{-270}$, so the \ion{O}{i} column density which we 
obtained in the LIC appears be higher than other evaluations. 

About \ion{N}{i}, Meyer et al.~(\cite{meyer_NI}) derived the average 
ISM abundance $10^6 \times N($\ion{N}{i}$)/N($\ion{H}{i}$)=75\pm4$, 
which is more accurate than the one obtained by Ferlet~(\cite{ferlet81}): 
$10^6 \times N($\ion{N}{i}$)/N($\ion{H}{i}$)=62^{+45}_{-34}$. We obtained 
inside BC and LIC $37\pm15$ and $32\pm7$ respectively, which appear to 
be low values, although compatible with the Ferlet~(\cite{ferlet81}) 
value. We note that already Vidal-Madjar et al.~(\cite{avm98}) found 
a low  \ion{N}{i} abundance on average toward G191-B2B: 
$10^6 \times N($\ion{N}{i}$)/N($\ion{H}{i}$)=33\pm2$. 
This seems to confirm a rather low \ion{N}{i} content in the 
local ISM relative to the average ISM value. 

On the other hand, column densities of \ion{C}{ii} and 
\ion{O}{i} in the LIC appear too high. These two lines being 
saturated and not clearly showing the two components as other 
metal lines, it could mean that the column densities derived 
for \ion{C}{ii} and \ion{O}{i} are slightly overestimated.

\begin{table}
\caption[]{Depletion factors relative to solar values (\ie\ solar 
abundance divided by measured interstellar abundance). Solar 
abundances are from Anders \& Grevesse~(\cite{ag89}).}
\label{depletion}
\begin{tabular}{l|cc||l|cc}
\hline
 & BC & LIC & & BC & LIC \\
\hline
\ion{N}{i}/\ion{H}{i}   & 2.6  & 4.2  & \ \ \ \ \ -- & -- & -- \\
\ion{O}{i}/\ion{H}{i}   & 2.7  & 0.9 & \ion{O}{i}/\ion{N}{i} & 1.1  
& 0.2  \\
\ion{Si}{ii}/\ion{H}{i} & 2.3  & 5.0  &\ion{Si}{ii}/\ion{N}{i} & 1.1  
& 1.2  \\
\ion{C}{ii}/\ion{H}{i}  & 1.2  & 0.4  &\ion{C}{ii}/\ion{N}{i}  & 0.5  
& 0.1  \\
\ion{Fe}{ii}/\ion{H}{i} & 19.0 & 13.8 &\ion{Fe}{ii}/\ion{N}{i} & 3.7  
& 3.3  \\
\ion{Mg}{ii}/\ion{H}{i} & 6.7  & 9.0  &\ion{Mg}{ii}/\ion{N}{i} & 2.6  
& 2.2 \\
\hline
\end{tabular}
\end{table}

\subsection{Ionization of the Local Interstellar Cloud}

\begin{figure}
\resizebox{\hsize}{!}{\includegraphics{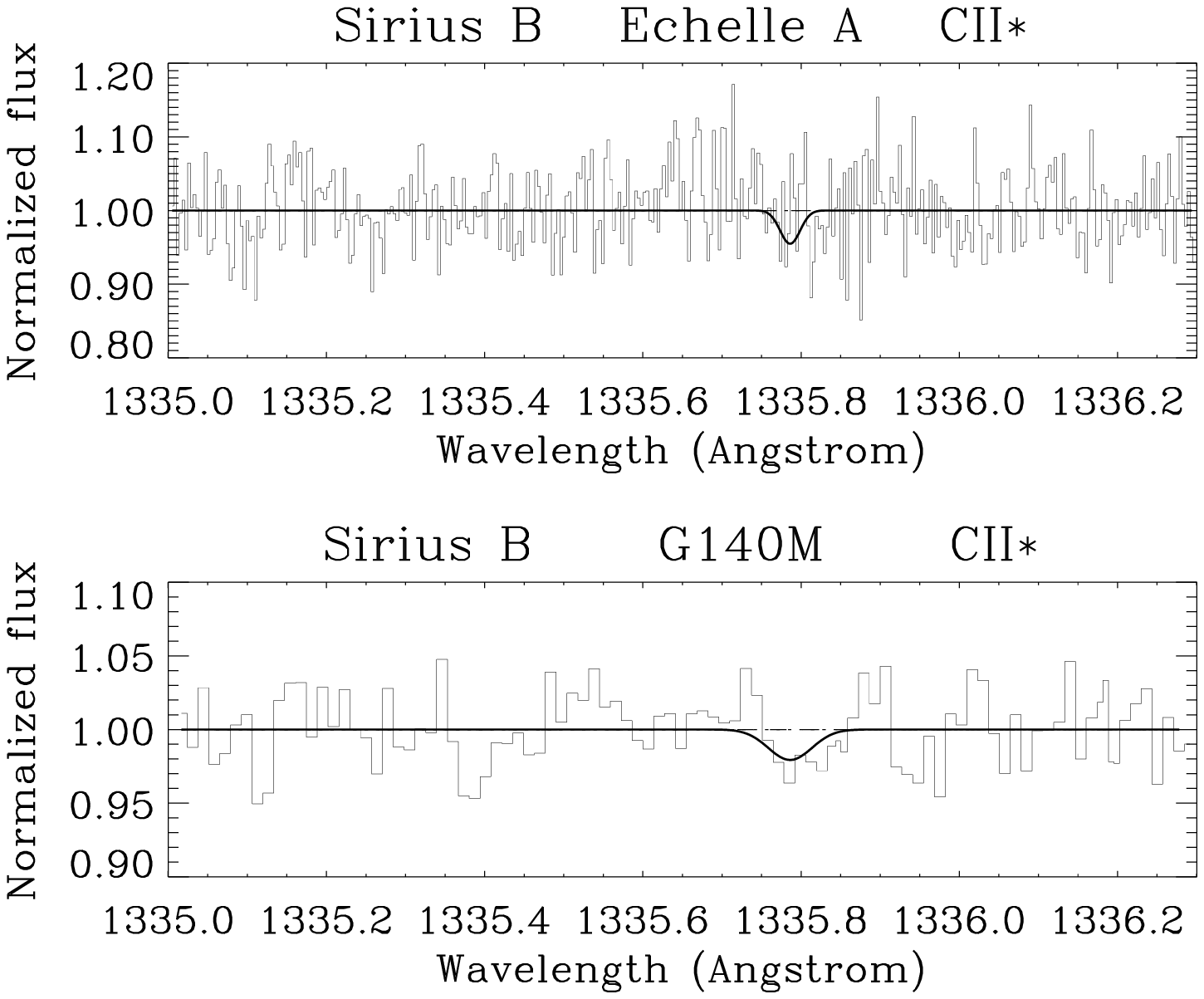}}      
\caption[]{High (top) and medium (bottom) resolution normalized spectra of 
the \ion{C}{ii*} line at 1335.7~\AA\ toward Sirius~B. The thick solid line 
corresponds to the limiting 1.5~m\AA\ detectable equivalent width at 
$3\sigma$. The interstellar line is not significantly detected.}
\label{cii_star}
\end{figure}

We obtained four spectra on the spectral region of the 1334~\AA\ 
\ion{C}{ii} line, toward Sirius~A or Sirius~B with high and 
medium spectral resolution. We did not significantly detect the 
interstellar excited-state \ion{C}{ii*} line at 1335.7~\AA. 
Assuming the above results for the width of the LIC lines, the 
criterion on the limiting detectable equivalent width at $3\sigma$ 
[$W_{lim}\equiv{3\Delta\lambda\over S/N}$, see 
H\'ebrard et al.~(\cite{heb})] 
gives the lower value $W_{lim}\simeq 1.5$~m\AA\ on Sirius~B spectra
(see Fig.~\ref{cii_star}) and the upper limit of 
$N_{\mathrm{LIC}}$(\ion{C}{ii*})$~\le 8\times 10^{11}$ cm$^{-2}$
[assuming $f_{\mathrm{\ion{C}{ii*}}}=0.1149$ (Morton~\cite{morton})]. 
We thus obtained: 
$$\left[\frac{N(\mathrm{\ion{C}{ii*}})}{N(\mathrm{\ion{C}{ii}})}\right]_{\mathrm{LIC}}\le 3 \times 10^{-3}.$$

Wood~\& Linsky~(\cite{wl97}) used this ratio to derive the electron 
density from the equation of the equilibrium between collisional 
excitation to excited-state \ion{C}{ii*} and radiative de-excitation 
to ground-state \ion{C}{ii}:
$$N(\ion{C}{ii*}) \times A_{21} = N(\ion{C}{ii}) \times n_e \, C_{12}(T).$$
These authors derive the value $n_e=0.11^{+0.12}_{-0.06}$~\cmm3\ for 
the LIC toward Capella. Using the same method toward the white dwarf 
REJ~1032-532, Holberg et al.~(\cite{holberg99}) derive the same value 
for the LIC: $n_e=0.11^{+0.07}_{-0.06}$~\cmm3 . Our non-detection of 
\ion{C}{ii*} in the LIC toward Sirius gives the limit $n_e\le0.05$~\cmm3, 
following Wood~\& Linsky~(\cite{wl97}) and references therein in order 
to find the value for the radiative de-excitation rate coefficient 
$A_{21}$ and determine the value of the collision rate coefficient 
$C_{12}(T)$ for $T$=8000~K. Wood~\& Linsky~(\cite{wl97}) used the 
value $T=$7000~K, which is the temperature usually found for the LIC. 
Using $T=$7000~K in our calculation leads to a very slightly and 
non-significantly lower value for $n_e$ compared to the one 
obtained using $T=$8000~K. 

\begin{figure}
\resizebox{\hsize}{!}{\includegraphics{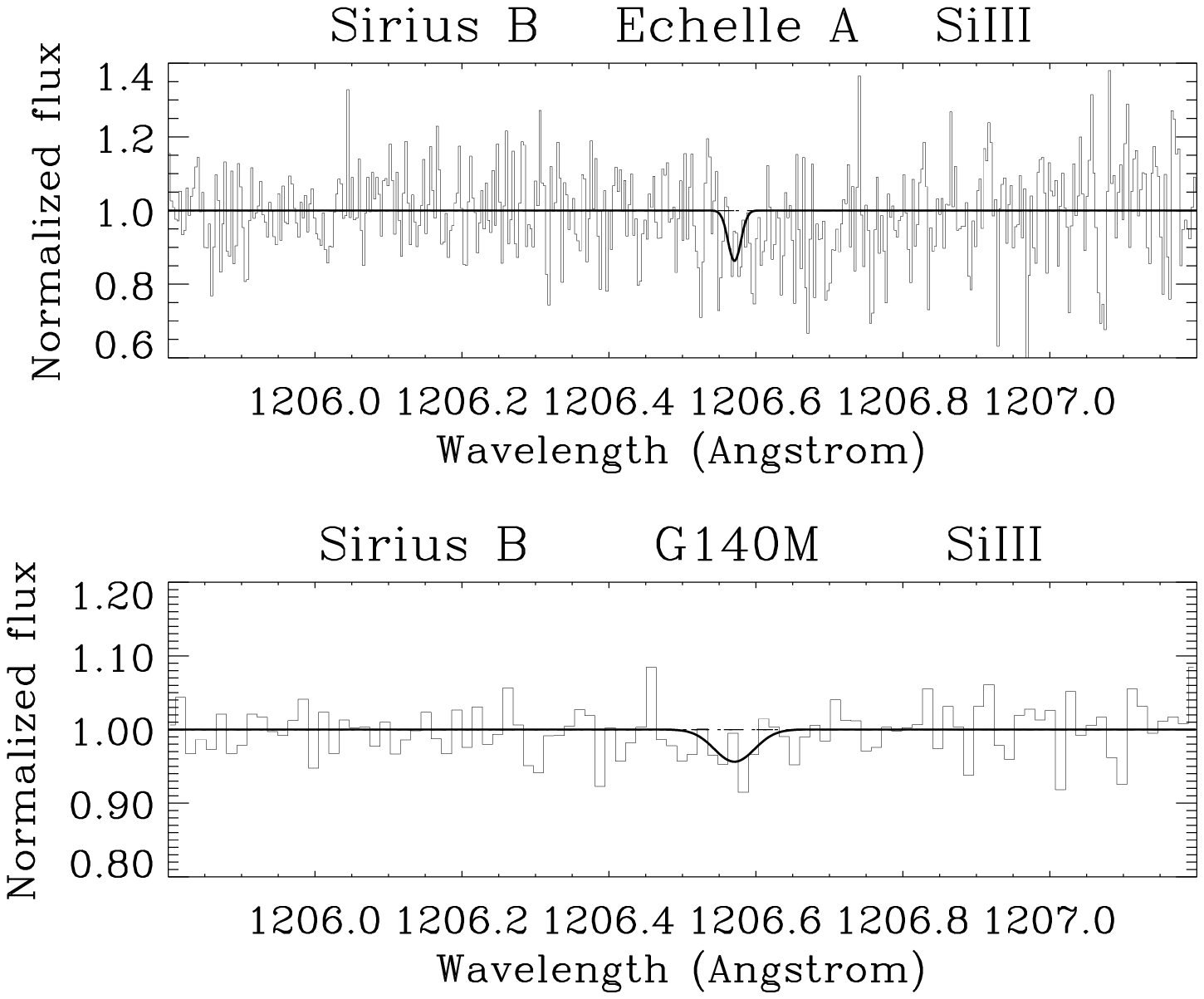}}      
\caption[]{High (top) and medium (bottom) resolution normalized spectra 
of the \ion{Si}{iii} line at 1206.5~\AA\ toward Sirius~B. The thick 
solid line corresponds to the limiting 3~m\AA\ detectable equivalent 
width at $3\sigma$. The interstellar line is not significantly detected.}
\label{siiii}
\end{figure}

This low value of $n_e$ indicates a rather low ionization, which could be 
argued for also by the non-detection of the \ion{Si}{iii} interstellar 
line at 1206.5~\AA. The Sirius~A \ion{Si}{iii} photospheric line appears 
near 0~\kms\ but no interstellar \ion{Si}{iii} line is significantly 
detected on any of our four spectra in that spectral region, toward 
Sirius~A or Sirius~B with high and medium spectral resolution. The lower 
limiting detectable equivalent width at $3\sigma$ found, obtained on 
Sirius~B spectra (see Fig.~\ref{siiii}), is $W_{lim}\simeq 3$~m\AA, from 
which we derive the upper limit 
$N_{\mathrm{LIC}}$(\ion{Si}{iii})~$\le 1.5\times 10^{11}$~cm$^{-2}$
[assuming $f_{\mathrm{\ion{Si}{iii}}}=1.669$ (Morton~\cite{morton})].
We thus obtained: 
$$\left[\frac{N(\mathrm{\ion{Si}{iii}})}{N(\mathrm{\ion{Si}{ii}})}\right]_{\mathrm{LIC}}\le 6 \times 10^{-2}.$$

This non-detection of the line \ion{Si}{iii} at 1206.5~\AA\ is surprising 
because this line was detected in the LIC by Gry et al.~(\cite{gry95}) 
toward $\epsilon$~CMa which is located only 12\degr\ away from Sirius. 
They derived the column density 
$N_{\mathrm{LIC}}($\ion{Si}{iii}$)=2.0\pm0.2 \times 10^{12}$~\cm2\ 
which is a value more than 10 times larger than our upper limit! 
If the line detected by Gry et al.~(\cite{gry95}) is actually 
caused by the LIC and not by another more distant cloud whose radial 
velocity is by coincidence confused with the LIC one, this may mean 
that the ionization in the LIC varies over very short distances. 
This non-detection may confirm the non-detection of \ion{Si}{iii} 
in the LIC already reported toward G191-B2B by 
Vidal-Madjar et al.~(\cite{avm98}). 
Moreover, Holberg et al.~(\cite{holberg99}) claimed that it is quite 
doubtful that the \ion{Si}{iii} ISM absorption line that they detected 
toward REJ~1032-532 is produced in the LIC. 
Thus it seems to favour the idea that another cloud in the long line 
of sight toward $\epsilon$~CMa is by coincidence at a radial velocity 
similar to the LIC one. 
Note that Gry et al.~(\cite{gry95}) do not detected \ion{Si}{iii} in 
their ``component~2'', in agreement with our non-detection of that ion 
in ``BC'' toward Sirius, which is identified as the same component. 

In addition, Dupin \& Gry~(\cite{dupin98}) detected a saturated 
\ion{Si}{iii} line at 1206.5~\AA\ in their component ``D'' toward 
$\beta$~CMa, which is located at less than 6\degr\ from Sirius. They 
derived the column 
density $N($\ion{Si}{iii}$)=(1.5 - 10) \times 10^{14}$~\cm2 . Since 
the velocity of that component does not correspond to the LIC, this 
huge \ion{Si}{iii} column density could be interpreted as a sign that 
the component ``D'', detected by Dupin \& Gry~(\cite{dupin98}) toward 
$\beta$~CMa, is probably located beyond Sirius and may explain some 
\ion{Si}{iii} absorption in the $\epsilon$~CMa line of sight, for 
instance at the LIC velocity.

\section{Discussion}
\label{Discussion}

Our \dsh\ evaluation is made assuming the three following main hypotheses:

\begin{itemize}
\item
The structure of the interstellar medium is the same toward Sirius~A and 
Sirius~B. Thus we detect the same interstellar clouds with the same 
physical properties toward these two targets.
\item
The \ion{H}{i} and \ion{D}{i} interstellar structure at \lya\ is the one 
determined through the \ion{O}{i}, \ion{N}{i}, \ion{Si}{ii}, \ion{C}{ii}, 
\ion{Fe}{ii} and \ion{Mg}{ii} lines. Then the physical properties 
(temperature, turbulence, velocity) at \lya\ are the same as in the metal 
lines, and there is no extra \ion{H}{i} or \ion{D}{i} interstellar 
component in addition to the BC and the LIC.
\item 
The processes which are responsible for the extra absorption in 
the blue wing of the \lya\ line toward Sirius~A and in the red 
wing of the \lya\ line toward Sirius~B are from stellar origin 
and do not perturb the BC and LIC absorptions. In addition, 
each process on each \lya\ wing does not reach the other wing. 
Thus, the composite \lya\ line formed by the red wing of the 
Sirius~A and the blue wing of the Sirius~B is a ``pure interstellar'' 
\lya\ line, and presents absorptions caused only by the BC and LIC.
\end{itemize}

The first assumption is justified by the small distance between the two 
stars ($\sim10$~AU). The second assumption is less strongly justified 
and one can imagine a low column density \ion{H}{i} component without 
detectable metal line, but with a still significant absorption at \lya. 
The most we can argue is that it is possible to fit the ``pure 
interstellar'' \lya\ line only with the BC and the LIC as 
found with the metal lines and that no extra component is 
required by the data. We did not find any width or 
velocity structure unexplained by our simple interstellar model 
(\ie\ only two components), as Linsky \& Wood~(\cite{linsky96}) 
did \eg\ on the sightline of $\alpha$~Cen. This led them to the 
detection of an extra \ion{H}{i} component they called the 
``hydrogen wall''. We are however unable to formally exclude  
the presence of additional low \ion{H}{i} column density clouds. 
The third assumption is justified 
again by the proximity between the two targets whereas their \lya\ 
absorptions are very different. This favours processes linked to the 
stars as causes of the extra \lya\ absorptions. Finally these 
suspected processes, the wind from Sirius~A and the Sirius~B photospheric 
line shape, are unable to disturb the other wing of the \lya\ line.

The reliability of our result is linked to the robustness of these 
hypotheses and to the possible inaccuracy in the metal column densities, 
as described in the \S~\ref{abundances_and_depletions}. We can note 
however that the \ion{C}{ii} and \ion{O}{i} column densities, suspected 
to be too high, do not seriously constrain the \lya\ fit. Indeed, the 
radial velocity shift of 5.9~\kms\ between BC and LIC applied to the 
\lya\ line is essentially constrained by the \ion{Fe}{ii}, \ion{Mg}{ii} 
and \ion{N}{i} lines which only show clearly the BC and LIC components.
Moreover, we used \ion{N}{i} and not \ion{O}{i} as tracer of \ion{H}{i}, 
in order to argue that the \ion{H}{i} column density ratio 
between LIC and BC should be probably ranging between 0.5 and~4. We thus 
concluded that possible inaccuracies in the \ion{C}{ii} and \ion{O}{i} 
column densities do not affect our \dshism\ evaluation.

\ 

Although the value of \dshbc\ could be between 0 and $1.6\times10^{-5}$, 
the best result is obtained with a low value. We found thus a \dshism\ 
abundance which seems to be low in one of the two 
components (BC) toward Sirius without being able to find an artifact 
able to explain that result. O and N are not overabundant in this 
component. It is thus difficult to see BC as a cloud polluted by D free 
material ejected by the planetary nebula preceeding the formation 
of the white dwarf Sirius~B since furthermore 
its radial velocity shoud be blue-shifted, 
whereas the BC one is redshifted. This fact does not agree with the 
simple idea of BC being an expanding shell of material ejected by 
the planetary nebula related to the white dwarf Sirius~B.

After G191-B2B (Vidal-Madjar et al.~\cite{avm98}) and $\delta$~Ori 
(Jenkins et al.~\cite{jenkins99}) on the line of sight of which low 
\dshism\ were measured, Sirius seems to be a good candidate for 
finding another low interstellar deuterium abundance. 

The cause of these variations has to be understood in order to know 
what is the actual value of the \dshism , if any. It is difficult to 
see interstellar deuterium as the simple tracer of the galactic 
chemical evolution. The study of its possible variation as a function 
of the radial distance to the galactic center may help us in that matter. 

Moreover, if \dshism\ ratio actually presents dispersion, one can 
argue that the other deuterium abundance evaluations, \ie\ proto-solar 
and primordial abundances, can also present dispersion. Indeed, 
variations of interstellar abundance of deuterium was detected thanks 
to the large number of sightlines available (several tens), whereas 
proto-solar and primordial abundances are determined only from few 
targets. Taking into consideration this problem, observing 
deuterium in the interstellar medium toward a large number of 
sightlines is a good way to proceed. 

\section{Conclusions}

We have presented new spectroscopic observations of Sirius~A and 
Sirius~B performed using HST-GHRS. 14 interstellar lines were detected 
at high and/or medium spectral resolution. The sightline, which is 
assumed to present the same structure toward the two stars, is 
composed by two clouds: the Blue Component (BC) and the Local 
Interstellar Cloud (LIC), in agreement with the previous HST-GHRS 
observations of Sirius~A reported by Lallement et al.~(\cite{lalle}). 

The three main results of our observations are the following:

\begin{itemize}
\item
The \lya\ lines do not present the same profile toward Sirius~A and 
Sirius~B, an extra absorption being observed in the blue wing of the 
Sirius~A \lya\ line, and an extra absorption being observed in the red 
wing of the Sirius~B \lya\ line. We interpreted these excesses 
respectively as the signatures of the wind from Sirius~A and of the 
core of the Sirius~B photospheric \lya\ line.
\item
A composite \lya\ profile was constructed from these two lines and 
fitted in order to measure the \dshism\ ratio on the two components. 
Our data are compatible with \dsh\ ratio found in the LIC by 
Linsky et al.~(\cite{linsky93} \&~\cite{linsky95}), \ie\ 
\dshlic~$=1.60 \pm 0.09 ^{+0.05}_{-0.10}\times 10^{-5}$. 
Our result is \dshlic~$=1.6 \pm 0.4\times 10^{-5}$. 
In the other component, BC, we did not 
detected a significant \ion{D}{i} line. The ratio we derived is  
$0<$\dshbc$<1.6\times10^{-5}$.
\item
We did not detect the interstellar absorption of \ion{Si}{iii} at 
1206.5~\AA\ and \ion{C}{ii*} at 1335.7~\AA. This implies a low electron 
density $n_e$, for which we found the upper limit $n_e\le0.05$~\cmm3 in 
the LIC, assuming equilibrium between collisional excitation to 
excited-state \ion{C}{ii*} and radiative de-excitation to ground-state 
\ion{C}{ii}. Since measured values of the electron density are higher in 
the LIC toward other sightlines, the new value of $n_e$ toward Sirius 
could point to inhomogeneities in the Local Interstellar Cloud.
\end{itemize}

The data are thus consistent with \dshbc\ in the range 0 to 
$1.6\times10^{-5}$. The BC cloud is a candidate region for low \dshism, 
but no definite conclusion about \dsh\ can be made at this time.
We intend to continue the study of the Sirius system
until we can come to a definitive conclusion as to whether or not 
a low \dshism\ is present in the BC. In particular, it is critically 
important that this experiment be done again with deep
HST-STIS observations to study the \lya\ line and FUSE observations 
to study the higher Lyman lines. FUSE, the 
Far Ultraviolet Spectroscopic Explorer which was launched on 
June 24, 1999, will perform observations between 905 and 1187~\AA, 
at a spectral resolution of $R=\lambda/\Delta\lambda\simeq30,000$. 
The primary goal of that mission is to observe deuterium toward 
more than one hundred targets, from the Solar System and the local 
interstellar medium up to extragalactic low-redshift objects. 
These dedicated studies should greatly clarify the problem of the 
chemical evolution of deuterium.

\begin{acknowledgements}

We thank L.~Ben~Jaffel and J.-P.~Dumont for many stimulating discussions.
We also thank S.~Dreizler for the calculation of the NLTE profiles 
for Sirius~B, and P.~Couteau for informations about orbit of 
Sirius~B around Sirius~A. We would like to thank the STScI staff for 
rescheduling observations after an original failure. Especially, we are 
grateful to A.~Berman who noticed that the plan for the repeated 
observations would probably also fail, D.~Soderblom who proposed 
another approach that was successful, and A.~Schultz who implemented 
it. Finally we would thank the referee J.~Linsky for very useful 
and constructive comments.
D.K. acknowledges support by grants from the DLR.

\end{acknowledgements}

\end{document}